\documentstyle[11pt,aaspp4,epsf]{article}

\newbox\grsign \setbox\grsign=\hbox{$>$} \newdimen\grdimen \grdimen=\ht\grsign
\newbox\simlessbox \newbox\simgreatbox
\setbox\simgreatbox=\hbox{\raise.5ex\hbox{$>$}\llap
     {\lower.5ex\hbox{$\sim$}}}\ht1=\grdimen\dp1=0pt
\setbox\simlessbox=\hbox{\raise.5ex\hbox{$<$}\llap
     {\lower.5ex\hbox{$\sim$}}}\ht2=\grdimen\dp2=0pt
\def\simgreat{\mathrel{\copy\simgreatbox}}
\def\simless{\mathrel{\copy\simlessbox}}
\def\vol#1  {{{#1}{\rm,}\ }}

\def\aj{{AJ}, }  
\def\apj{{ApJ}, } 
\def\apjs{{ApJS}, } 
\def\pasp{{PASP}, }  
\def\mnras{{MNRAS}, } 
\def\nat{{Nat}, }     
\def\aa{{A\&A}, }     

\begin{document}
\title{A Group--Group Merger at a Redshift of $z = 0.84$?}

\author{Lori M. Lubin\altaffilmark{1}}
\affil{Palomar Observatory, California Institute of Technology, Mail Stop 105-24, Pasadena, CA 91125}
\affil{lml@astro.caltech.edu}

\author{Marc Postman}
\affil{Space Telescope Science Institute\altaffilmark{2}, 3700 San Martin Drive, Baltimore, MD 21218}
\affil{postman@stsci.edu}

\author{J. B. Oke}
\affil{Palomar Observatory, California Institute of Technology, Pasadena,
CA 91125}
\affil{and}
\affil{National Research Council Canada, Herzberg Institute of Astrophysics, 
Dominion Astrophysical Observatory, 5071 W. Saanich Road, Victoria,
BC V8X 4M6}
\affil{oke@dao.nrc.ca}

\vskip 1 cm
\centerline{Accepted for publication in the {\it Astronomical Journal}}

\altaffiltext{1}{Hubble Fellow}

\altaffiltext{2}{Space Telescope Science Institute is operated by the
Association of Universities for Research in Astronomy, Inc.,
under contract to the National Aeronautics and Space Administration.}

\vfill
\eject

\begin{abstract}

We present a dynamical study of the CL0023+0423 system at a redshift
of $z = 0.84$.  This system consists of two components separated in
velocity space by $\sim 2900~{\rm km~s^{-1}}$ and on the plane of the
sky by $\sim 0.23~h^{-1}~{\rm Mpc}$. A kinematic analysis indicates
that the two components are a poor cluster with a velocity dispersion
of $415^{+102}_{-63}~{\rm km~s^{-1}}$ and a mass of $\sim 3 - 6 \times
10^{14}~h^{-1}~{\rm M_{\odot}}$ and a less massive group with a
velocity dispersion of $158^{+42}_{-33}~{\rm km~s^{-1}}$ and a mass of
$\sim 10^{13}~h^{-1}~{\rm M_{\odot}}$ (Postman, Lubin \& Oke
1998). The dynamics of galaxy groups at high redshift can provide
important insights into the creation of present-day galaxy clusters.
Therefore, we have performed a dynamical study on this system in order
to determine whether the two groups are infalling. This analysis
includes an analytic two-body calculation and N-body simulations.  The
results of both studies indicate that the system is most likely not
bound but simply a chance projection on the sky; however, within the
observational uncertainties, there do exist bound solutions where the
two galaxy groups are currently moving toward each other and will
eventually merge into a larger system of galaxies.  We have run
one-thousand N-body simulations with random initial conditions based
on the observed parameters of the CL0023+0423 groups. A statistical
analysis of these simulations indicates that there is an 20\% chance
that the two groups will merge.  If the CL0023+0423 system does merge,
it will appear as a cluster on the sky, as well as in velocity space,
within $1-2$ Gyrs. The cluster will evolve dynamically for more than 3
Gyrs, appearing during this time more similar to an open, irregular
cluster. The final merged system has a velocity dispersion which is
consistent with a local Abell richness class 1 cluster.

The morphological analysis of the galaxy populations of CL0023+0423
suggests that both groups are largely dominated by spiral galaxies.
Early-type fractions are 33\% or less (Lubin et al.\ 1998).  These
modest early-type fractions have implications for both cluster
formation and group evolution. Studies of open clusters at $z = 0.31 -
0.54$ indicate that that they have early-type fractions between $45 -
80\%$ (Dressler et al.\ 1997; Andreon, Davoust \& Helm 1997; Stanford,
Eisenhardt \& Dickinson 1997; Couch et al.\ 1998).  If the CL0023+0423
system is the predecessor of such a cluster, the comparison may
suggest that some fraction of early-type galaxies are formed between
redshifts of $z \sim 0.8$ and $z \sim 0.3$; however, the morphological
fractions are still highly uncertain.  In addition, the modest
early-type fractions in both groups may be inconsistent with the
strong correlation between velocity dispersion and early-type fraction
observed in nearby groups of galaxies (Zabludoff \& Mulchaey 1998).
Both groups apparently have relatively low early-type populations,
irrespective of their velocity dispersion.  If the groups of
CL0023+0423 are typical of galaxy groups at high redshift, and if
high-redshift groups are the progenitors of local groups, this result
may also imply that some early-type formation is occurring at
redshifts of $z \simless 0.8$.  These results do not preclude the
formation of early-type galaxies at very high redshift as many
observations suggest (e.g.\ Steidel et al.\ 1996; Ellis et al.\ 1997;
Stanford, Eisenhardt \& Dickinson 1997; Postman, Lubin \& Oke 1998);
however, the observations of the CL0023+0423 system may imply that a
fraction of the early-type population is forming and/or undergoing
significant evolution at redshifts of $z < 1$.
\end{abstract}


\section{Introduction}

In hierarchical clustering models, clusters of galaxies are formed by
a progressive coalescence of an inhomogeneous system of subclusters
(e.g.\ White 1976; Evrard 1990; Lacey \& Cole 1993,1994).  The early
stages of these models provide a natural description of the formation
of open, irregular clusters like Virgo or Hercules. These systems show
no single, central condensation, though the galaxy surface density is
at least five times as great as the surrounding field.  They often
have high degrees of asymmetry and significant amounts of
subclustering.  The progressive stages of evolution in cluster
formation produces centrally-concentrated, compact clusters of
galaxies. These clusters are dense, have a single, prominent
concentration among the bright member galaxies, and typically display
a high-degree of spherical symmetry (Abell 1958; Oemler 1975; Bahcall
1975; Dressler 1978; Dressler 1980a,b).

The remnants of hierarchical structure formation can be easily
observed in the local universe. Substructure and more significant
merger events are common characteristics of the optical and X-ray
properties of nearby clusters of galaxies (e.g.\ Geller \& Beers 1982;
Dressler \& Shectman 1988; West \& Bothun 1990; Jones \& Forman 1992;
Mohr, Fabricant \& Geller 1993; Zabludoff \& Zaritsky 1995).  These
studies indicate that $30 - 40\%$ of all clusters have substantial
substructure where the subclump contains a significant fraction of the
cluster galaxies (Dressler \& Shectman 1987; West \& Bothun 1990;
Jones \& Forman 1990). Simulations have shown that, when a cluster
accretes smaller groups of galaxies, the effect of such a merger, such
as optical or X-ray morphologies, can last for $\sim 1$ Gyr (Evrard
1990). For a more significant merger of two equal-mass subclusters, it
takes more than 4 Gyr after the first encounter until the
collisionless dark matter (and galaxy) density contour shows a single
peak.  The exact time depends on the density and velocity distribution
of the dark matter.  A longer relaxation time is required when the
dark matter is spatially extended and/or when the velocity
distribution has a Gaussian shape (Nakamura, Hattori \& Mineshige
1995).  It is worthwhile to note that hydrodynamic simulations
indicate that the collisional, X-ray emitting intracluster gas can
relax on timescales much shorter than the dark matter (see e.g.\
Evrard 1990).

Spectral studies of rich clusters with significant substructure
indicate that $\sim 15\%$ of the early-type cluster members have signs
of current or on-going star formation (Caldwell \& Rose 1997).  These
observations may imply that merger activity can alter the star
formation histories of the early-type members. Starbursts may be the
result of shocks which are induced in the collision between the
intracluster medium of the cluster and that of the subcluster.
Zabludoff \& Mulchaey (1998) find that a similar fraction of the
early-type galaxies in groups of galaxies in the field have
experienced star formation within the last $\sim 2$ Gyr. Therefore, if
some of the subclusters in these rich clusters are actually groups
which have fallen into the cluster environment, the similarity between
the star formation histories of the early-type galaxies in the
subclusters and those in the galaxy groups indicate that the group
environment may be the site of recent star formation. The modest
intragroup medium and low velocity dispersions would argue that
galaxy-galaxy collisions would be the dominant means to alter the star
formation history of the early-type galaxies in these systems (Aarseth
\& Fall 1980; Barnes 1985; Merritt 1995; Zabludoff \& Mulchaey 1998).
The results suggest that both group and cluster environments should be
examined as important sites for evolution in the star formation
history of galaxies.

In order to understand the evolution of galaxies and their
environments, it is essential to study groups and clusters of galaxies
in their early stages of formation.  Unfortunately, because of the
substantial background contamination rates at high redshift, finding
systems which are in the process of forming is very difficult.
However, we have found such a system as part of a high-redshift
cluster survey.  At a redshift of $z = 0.84$, the CL0023+0423 system
is part of an extensive photometric, spectroscopic, and morphological
study of nine candidate clusters of galaxies at $z \simgreat 0.6$
(Oke, Postman \& Lubin 1998, hereafter Paper I).  First detected as a
likely cluster by Gunn, Hoessel \& Oke (1986), more detailed
spectroscopic data reveal that this system is actually made up of two
substructures separated closely in velocity, as well as position on
the sky. The radial velocity histograms of the individual structures
more closely resemble local groups or poor clusters of galaxies
(Postman, Lubin \& Oke 1998, hereafter Paper II). In addition, the
morphological analysis of the galaxy population in this system
indicates that these groups do not have a morphological composition
which is typical of a massive, relaxed cluster. Their galaxy
population is almost totally dominated by spiral galaxies, with
early-type fractions that are 33\% or less (Lubin et al.\ 1998,
hereafter Paper III). This study provides the first observational data
on a possible group-group merger at high redshift.

Because of the intriguing nature of this system and its potential to
reveal important information on the hierarchical formation of larger
systems such as clusters of galaxies, we have decided to study in
detail the dynamical state of this two group system. In this paper, we
present the results of an analytic analysis of the system dynamics, as
well as N-body simulations of the group-group system.  We address the
possibility that this system is bound and will eventually merge, or
that the system is simply a chance projection of two groups on the
sky. A brief description of the observations are given in Sect.\ 2. In
Sects.\ 3 and 4, we describe the analytic dynamical analysis and the
N-body simulations, respectively. A discussion of the results and
their implications for the formation and evolution of groups and
clusters of galaxies is presented in Sect.\ 5.  In the following
analyses, we have assumed $q_{0} = 0.1$ and $H_{0} = 100~h~{\rm
km~s^{-1}~Mpc^{-1}}$.

\section{The Observations}
 
As part of our observational program on high-redshift clusters of
galaxies, the CL0023+0423 system has been observationally well
studied.  The details of the Keck optical observations, both broad
band and spectroscopic, and the HST imagery are presented in Papers I,
II and III.  In this section, we discuss briefly only those
observational parameters which are applicable to the following
dynamical analyses.

\subsection{Velocity Histograms and Group Masses}

\begin{figure}
\centerline{\epsfbox{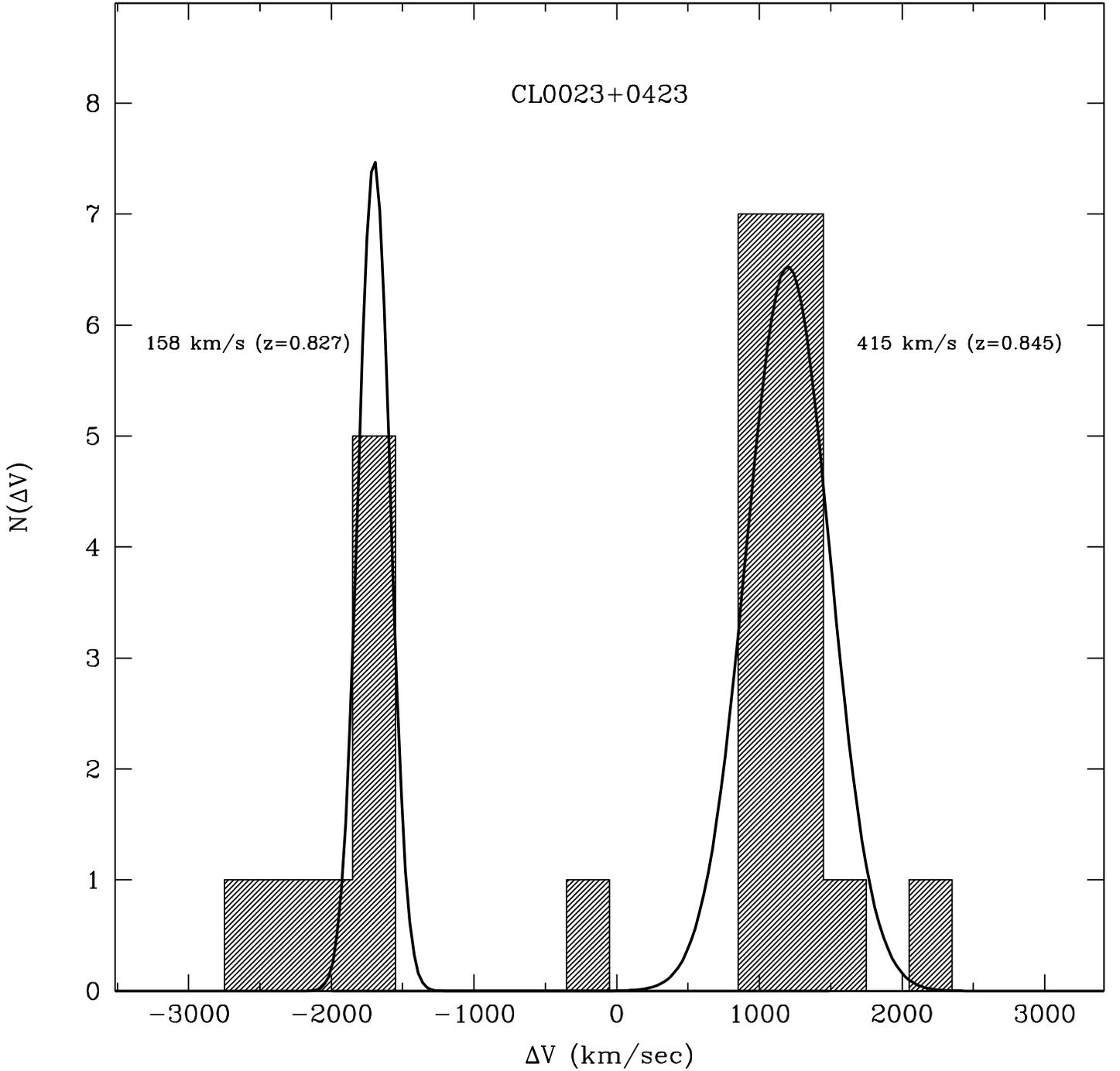}}
\caption{Histogram of the relativistically corrected radial velocity offsets
for CL0023+0423. The best-fit Gaussian distributions are shown for
comparison. The low-velocity component CL0023A has a mean redshift of
$\bar{z} = 0.827$ and a velocity dispersion of $158^{+42}_{-33}~{\rm
km~s^{-1}}$.  The high-velocity component CL0023B has a mean redshift
of $\bar{z} = 0.845$ and a velocity dispersion of
$415^{+102}_{-63}~{\rm km~s^{-1}}$.}
\label{velhist}
\end{figure}

From the observations described above, we have analyzed the kinematic
properties of the CL0023+0423 system.  The procedure for computing the
cosmologically and relativistically corrected velocity histograms are
discussed in Sect.\ 3 of Paper II. The resulting velocity distribution
for this system is shown in Figure~\ref{velhist}. It is clear from
this figure that there is a clear bimodal distribution in redshift
space, corresponding to peaks at $z = 0.8274$ (7 galaxies) and $z =
0.8452$ (17 galaxies). This implies a cosmologically-corrected radial
velocity difference of $V_r = 2922 \pm 216~{\rm km~s^{-1}}$.
Following the convention of Paper II, we refer to the low-velocity
system as CL0023A and the high-velocity system as CL0023B. The
velocity dispersions of the individual components are
$158^{+42}_{-33}$ and $415^{+102}_{-63}~{\rm km~s^{-1}}$ for CL0023A
and CL0023B, respectively.  The uncertainty in the dispersion is
computed following the prescription of Danese et al.\ (1980). The
Danese et al.\ prescription assumes the errors in velocity dispersions
can be modeled as a $\chi^2$ distribution and that the velocity
deviation from the mean cluster redshift is independent of galaxy mass
(that is, the cluster is virialized).  For more details on the
velocity dispersion measurements, see Paper II.  We have also
calculated the harmonic radius ($R_h$) for each system (Equation 2 of
Paper II). These values are $0.116$ and $0.366~h^{-1}~{\rm Mpc}$ for
CL0023A and CL0023B, respectively. Following the formalism in Ramella,
Geller \& Huchra (1989), the radius and velocity dispersion of each
system imply virial crossing times (in units of the Hubble time) of
0.02. Such short crossing times are consistent with other bound
systems of galaxies, such as groups and rich clusters (Ramella, Geller
\& Huchra 1989).

We have used three separate estimators to determine the masses of
virialized systems given a set of galaxy positions and
redshifts. These estimators include the pairwise mass ($M_{PW}$), the
projected mass ($M_{PM}$), and the ringwise mass estimators
($M_{RW}$). The exact formalisms are described in detail in Sect.\ 3.1
of Paper II. The differences in the mass estimates are due primarily
(but not solely) to the difference in the radius estimators. The
pairwise estimator gives a high weight to close pairs; the other two
estimates are less sensitive to this and, therefore, give more similar
values (see Paper II and references therein).  For the following
analyses, we have chosen the ringwise mass estimate as the actual mass
of each structure (Carlberg et al.\ 1996). This corresponds to a total
system mass of $M = 5.94 \pm 0.91 \times 10^{14}~{\rm M_{\odot}}$. The
velocity dispersion, the harmonic radius, and all three masses of each
system are listed in Table~\ref{table-mass}.  All of these parameters
are very similar to those of local galaxy groups and poor (Abell
richness class 0) clusters (e.g.\ Huchra \& Geller 1982; Ramella,
Geller \& Huchra 1989; Zabludoff \& Mulchaey 1998 and references
therein). The number of confirmed members, the physical properties,
and the short crossing times of both CL0023A and CL0023B suggest that
these systems are real, physically-bound groups of galaxies.

\subsection{Group Separation}

\begin{figure}
\centerline{\epsfbox{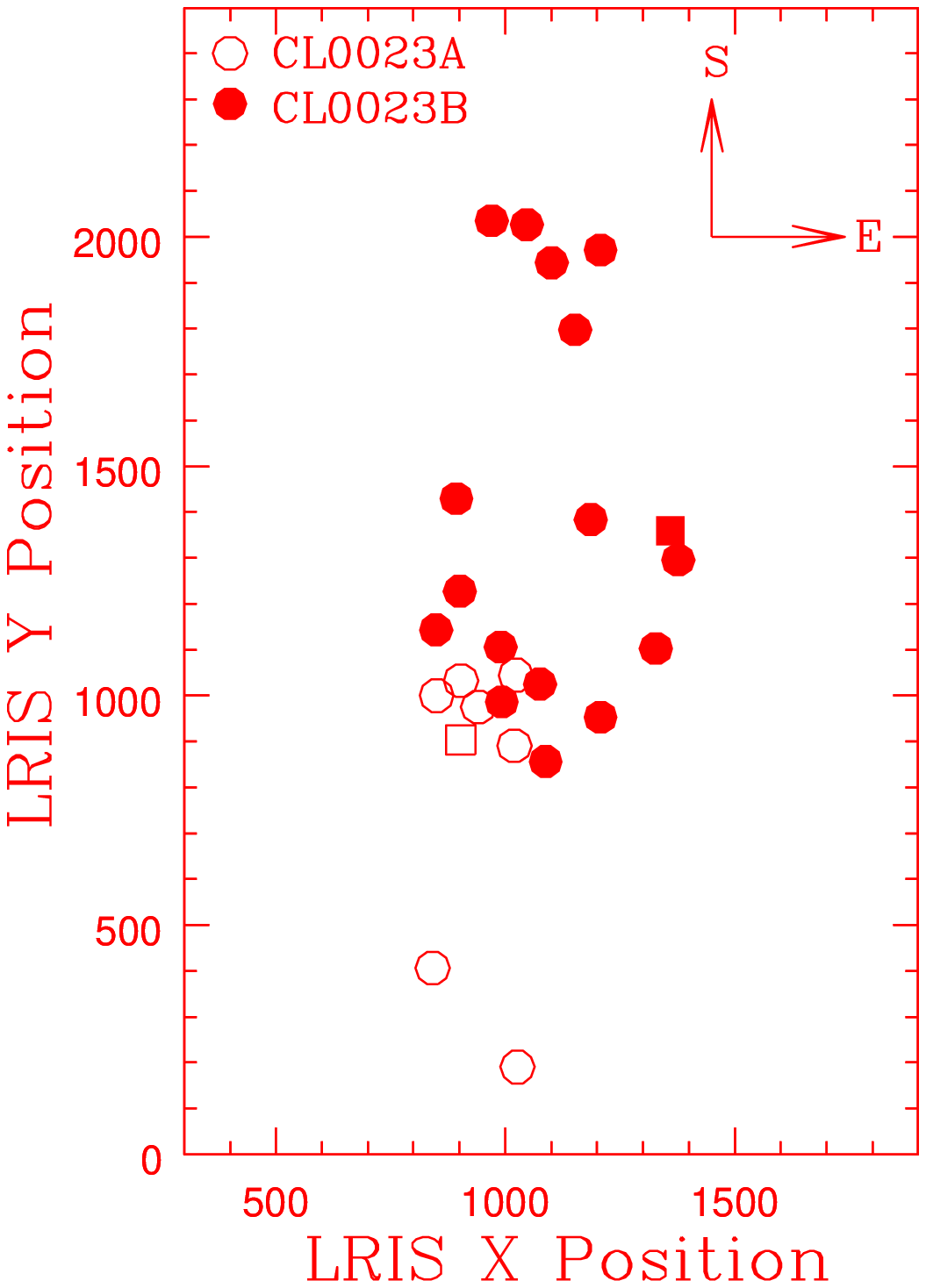}}
\caption{The LRIS XY positions of all confirmed members in the CL0023+0423
system. One unit on these axes is equivalent to 0.215 arcseconds.  The
galaxies associated with the low-velocity group CL0023A are indicated
by open symbols, while the galaxies associated with the high-velocity
group are indicated by closed symbols. The brightest member of each group
is indicated by a square. }
\label{xy}
\end{figure}

As further indication of the distinctiveness of each group, the two
velocity peaks are also separated on the sky (see Figure~\ref{xy}).
The groups do not have well-defined galaxy profiles.  Such galaxy
distributions are not uncommon for local groups of galaxies (see
Figure 3 of Zabludoff \& Mulchaey 1998).  However, because we need an
estimate of the projected separation between the two galaxy groups for
the dynamical analysis presented in Sect.\ 3, we must calculate the
centroid of CL0023A and CL0023B.  It is obvious from Figure~\ref{xy}
that more confirmed group members are necessary to get an accurate
measure of the group center (see Figure 4 of Paper II). Because of
this uncertainty, we have estimated the projected separation in two
ways.  In both methods we have used only the positions of the
confirmed members in each group.  Firstly, we have calculated the
separation between the {\it median} positions of each subsystem.  The
RA and Dec of the median positions of CL0023A and CL0023B are given in
Table~\ref{table-sep}. This projected separation between these two
positions is ${47}\farcs{2}$ or a physical separation of $R_p =
0.23~h^{-1}~{\rm Mpc}$.  Secondly, we have computed the separation
between the brightest group members.  The brightest confirmed member
in CL0023A is Keck \#2055 with an absolute AB magnitude of $B =
-21.63$ mag; the brightest confirmed member in CL0023B is Keck \#2348
with an absolute AB magnitude of $B = -21.49$ mag (see Paper II).  The
RA and Dec of these two galaxies are listed in Table~\ref{table-sep}.
The projected separation between their positions is ${103}\farcs{0}$
or a physical separation of $R_p = 0.50~h^{-1}~{\rm Mpc}$. Because of
the reasonably large disparity between the two calculated separations,
we perform an analytic analysis of the dynamics using each of the
projected separations.

\section{The Dynamics}

\subsection{The Analytic Analysis}

We examine three possible solutions : (1) the system is bound but
still expanding, (2) the system is collapsing, and (3) the two galaxy
groups are not bound to each other but merely a chance projection on
the sky.  We assume that, if the system is bound, that we are seeing
the system prior to complete merging and virialization.

We have analyzed the CL0023+0423 system based on the analytic
prescription used in Beers, Geller \& Huchra (1982) for Abell 98. The
system is considered a two-body problem with each galaxy group having
a linear orbit and being a point source with all of the mass
concentrated at the group center (see Sect.\ 2.2). The dynamical state
of the system is studied by solving the equations of motion as a
function of $\alpha$, the angle between the line connecting CL0023A
and CL0023B (see Figure~\ref{schem}).  We assume that at $t = 0$ the
two groups are at zero separation and that they are now coming
together or moving apart for the first time. If the system is still
expanding, the high-velocity component CL0023B must be more distant
than the low-velocity component CL0023A because the individual clumps
must have collapsed out of the Hubble flow.  If the cluster has
already reached maximum expansion and begun to collapse, CL0023B must
be nearer than CL0023A.

For the bound case, the parametric solutions to the equations of
motion are :

\begin{eqnarray}
R & = & \frac{R_m}{2} (1 - {\rm cos}\chi) \\ 
t & = & {\left(\frac{R_m^3}{8GM}\right)}^{1/2} (\chi - {\rm sin}\chi) \\
V & = & {\left(\frac{2GM}{R_m}\right)}^{1/2} \frac{({\rm sin}\chi)}{(1 - {\rm
cos}\chi)} 
\end{eqnarray}

\noindent where $R_m$ is the separation of the two groups at maximum
expansion, $M$ is the total mass of the system, and $\chi$ is the
developmental angle. 

For the unbound case, the parametric solutions to the equations of
motion are :

\begin{eqnarray}
R & = & \frac{GM}{V_{\infty}}({\rm cosh}\chi - 1) \\
t & = &	\frac{GM}{V_{\infty}^{3}}({\rm sinh}\chi - \chi) \\
V & = & V_{\infty} \frac{{\rm sinh}\chi}{({\rm cosh}\chi - 1)}
\end{eqnarray}

\noindent where $V_{\infty}$ is the asymptotic expansion velocity. The
observed parameters of radial velocity difference $V_{r}$ and the
projected separation $R_{p}$ are related to these equations by

\begin{eqnarray}
V_{r} & = & V~{\rm sin}\alpha \\
R_{p} & = & R~{\rm cos}\alpha
\end{eqnarray}

\noindent where $\alpha$ is the angle between the line connecting the
two groups and the $x$ axis (see Figure~\ref{schem}).

As described in Sect.\ 2.2, we will explore two possible projected
separations : case (1) $R_{p} = 0.23~h^{-1}~{\rm Mpc}$, the separation
between the median positions of each group; and case (2) $R_{p} =
0.50~h^{-1}~{\rm Mpc}$, the separation between the two brightest
galaxies in each group (see Table~\ref{table-sep}). We can solve these
systems of equations by constraining the time to be $t = t_o =
3.98~{\rm Gyr}$, the age of the universe at the mean redshift of the
two systems ($\bar{z} = 0.836$) for a $q_{o} = 0.1$ and $H_{0} =
100~h~{\rm km~s^{-1}~Mpc^{-1}}$ universe. In addition, the Newtonian
criterion for gravitational binding places a limit on the bound
solution from the observable parameters. That is,

\begin{equation}
V_{r}^{2} R_{p} \le 2 G M~{\rm sin}^{2}\alpha~{\rm cos}\alpha
\end{equation}

\noindent Using the above equations and constraints, we can solve the
equations of motion iteratively to find the projection angle $\alpha$
as a function of radial velocity difference $V_{r}$. Therefore, we
adopt the time condition of $t_o = 3.98~{\rm Gyr}$ and the observed
total mass of the system $M = 5.94 \times 10^{14} M_{\odot}$ (see
Table~\ref{table-mass}) to solve the equations of motion. The results
are plotted in Figure~\ref{av-act} which shows the $(\alpha, V_{r})$
plane for each value of $R_{p}$.  The shaded region specifies where
all unbound solutions would lie, while the unshaded region specifies
where all bound solutions would lie. The solutions for our particular
two-body problem are indicated by solid and dashed lines in this
figure.  The two solid lines indicate our bound solutions,
bound-incoming (BI) and bound-outgoing (BO), while the dashed line
indicates our unbound (UO) solution. It is obvious from this plot that
there are no bound solutions for this system. For case (1) of $R_p =
0.23~{\rm Mpc}$, there is in principle an incoming-bound solution for
the observed $V_r$ which does not violate energy conservation;
however, there are no formal incoming or outgoing-bound solutions
given the constraints of the equations of motion. For case (2) of $R_p
= 0.50~{\rm Mpc}$, the observed $V_r$ lies far from the bound region
of this plane, indicating that the two-body system in this
configuration is definitely unbound.

Because in case (1) the system is extremely close to having an
incoming-bound solution, we have examined the effect of varying the
two additional parameters, the total mass of the system $M$ and the
present age of the system $t_o$. Based on Equations 1--3, it is clear
that only more mass or a longer lifetime will make this system
bound. We examine both of these possibilities in Figure~\ref{av-plus}.
In the left panel, we show the $(\alpha, V_r)$ plane for a system
which has $1 \sigma$ more mass that the observed value, i.e.\ $M =
6.85 \times 10^{14}~M_{\odot}$ (see Sect.\ 2.1 and
Table~\ref{table-mass}); the projected separation $R_p$ and the age
$t_o$ remain the same at $0.23~{\rm Mpc}$ and $3.98~{\rm Gyr}$,
respectively.  In the right panel, we show the $(\alpha, V_r)$ plane
for a system which is twice as old with $t_o = 7.96~{\rm Gyr}$; the
projected separation $R_p$ and the total mass $M$ remain the same at
$0.23~{\rm Mpc}$ and $5.94 \times 10^{14}~{\rm M_{\odot}}$,
respectively.  For each of these cases, there are now bound solutions
within $\pm 1\sigma$ of the observed radial velocity difference $V_r =
2922 \pm 216~{\rm km~s^{-1}}$. Table~\ref{table-sol} lists the
possible range of values of the calculated parameters $\chi$,
$\alpha$, and $R_m$ for all bound-incoming solutions that give a
radial velocity difference that is within $1 \sigma$ of the observed
value of $V_r$.  We note that if the two groups are not virialized,
the masses derived from any estimator (see Sect.\ 2.1) will tend to be
overestimates (Small et al.\ 1998).  Hence, it may be unlikely that we
have underestimated the total mass of the system.  Only if the two
groups are fully virialized will a $+1\sigma$ deviation be as equally
plausible as a $-1\sigma$ deviation.

These inbound solutions imply that maximum expansion was achieved $2 -
4$ Gyrs ago.  The groups are now less than $0.5~{\rm Mpc}$ apart and
are moving at over $3500~{\rm km~s^{-1}}$ relative to each other. They
will cross ($R = 0$) within $\sim 5 \times 10^{7}$ years.  The
separation is so small and the relative velocity so high that the two
groups should already be experiencing significant dynamical
interactions and be well into the process of merging.

\section{N-Body Simulations}

In addition to the analytic calculations, we have used an N-body tree
code to simulate the time evolution of our group-group system. This
code is described in Barnes \& Hut (1986,1989) and is available
publically at the website
ftp://hubble.ifa.hawaii.edu/pub/barnes/treecode/.  Hierarchical tree
codes are based on the concept of approximating the long-range force
field of a localized region with a multipole expansion.  A
tree-structure partitions the system into a hierarchy of such
regions. The gravitational field at any point is approximated by
making a partial recursive descent of the tree to obtain a detailed
description of the nearby mass distribution and a coarser description
of the more distant parts of the system (see also Hernquist 1987). If
a better approximation is necessary, the algorithm may either examine
a finer level of the hierarchy or include more moments of the mass
distribution for each region.

The usefulness of the N-body simulations is that we are able to run
many simulations with a variety of initial conditions.  The initial
conditions relating to the particles themselves include their total
number, the mass of each particle, and the three-dimensional position
and velocity of each particle.  We use the current observational
configuration of the CL0023+0423 system to create the initial setup of
our simulations. We begin by creating two groups of particles which we
refer to as Group A and Group B.  These two systems are analogous to
the actual groups CL0023A and CL0023B and have properties based on the
observed values.  Because the mass ratio of CL0023B to CL0023A is
approximately 17:1 (see Table~\ref{table-mass}), we include 17 times
more particles in Group B; this ensures that each particle in the
system has roughly equal mass. We include 100 particles in Group A in
order to have a reasonable statistical sample for this group, as well
as a sufficient number of total particles (1800).  The specific
numbers and ratios of particles do not seriously affect the
statistical outcome of the simulations (see below).  The exact
particle mass is determined by choosing the total mass of each group
and dividing by the respective number of particles in that group.  In
order to explore a wide range of parameter space, we randomly choose a
mass of each group that is within $\pm 2 \sigma$ of the observed value
(see Table~\ref{table-mass}).

\begin{figure}
\epsfysize=7.0in
\centerline{\epsfbox{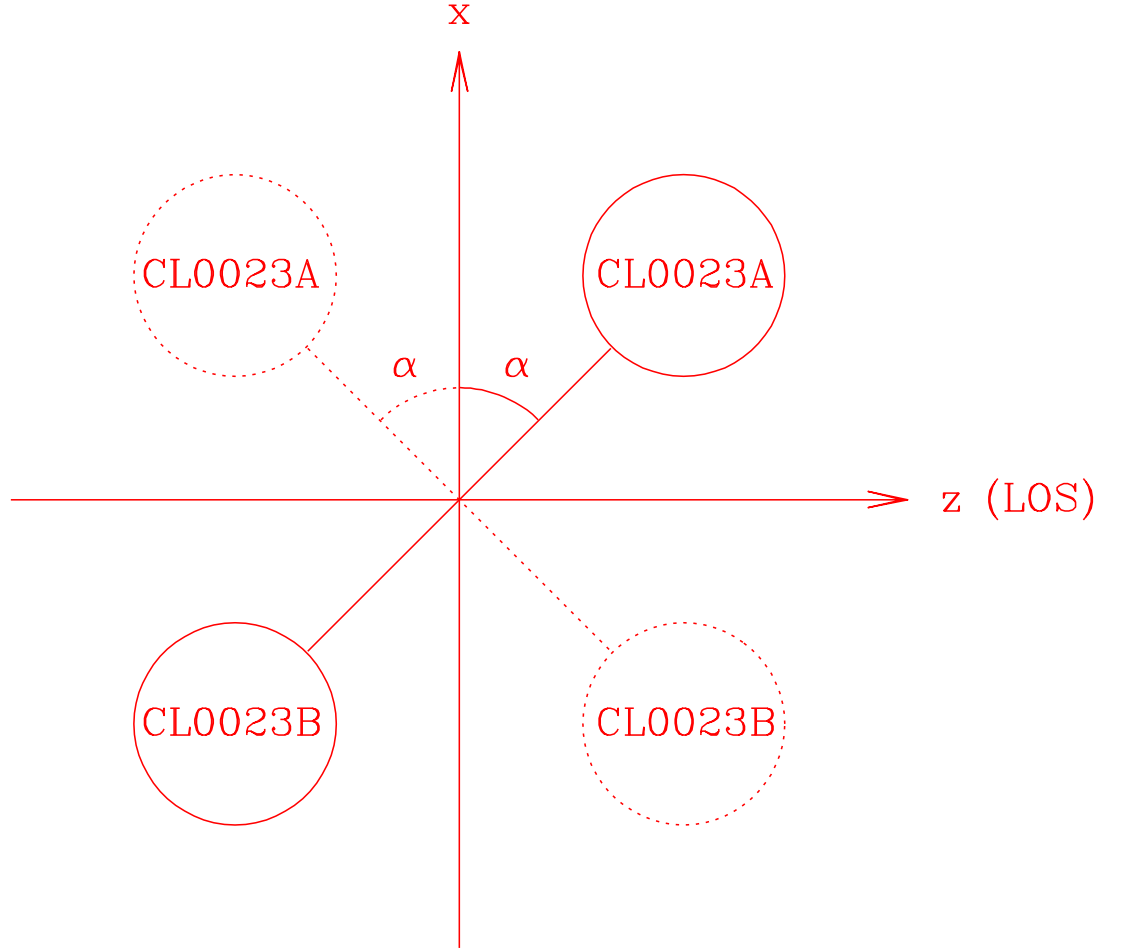}}
\caption{The physical schematic of the CL0023+0423 system. The case
where the system is still expanding is indicated by the solid
lines. In this case, the high velocity component CL0023B must be more
distant than the low velocity component CL0023A because the individual
clumps must have collapsed out of the Hubble flow. The case where the
system has already reached maximum expansion and begun to collapse is
indicated by the dotted lines. In this case, the CL0023B must be
nearer than CL0023A. $\alpha$ represents the angle between the line
connecting the two groups CL0023A and CL0023B. The $x$ and $z$
(line-of-sight) axes are indicated. The $y$ axis is coming out of the
page (see Sect.\ 4). }
\label{schem}
\end{figure}

We must now place each of these particles in three-dimensional space.
The adopted coordinate system is that shown in Figure~\ref{schem}
where $x$ is the transverse direction, $y$ is the vertical direction
(coming out of the page), and $z$ is the direction along the
line-of-sight. The configuration of the two group system is taken to
be that in a bound-incoming solution (the dotted line in
Figure~\ref{schem}).  We have shown in Sect.\ 3 that, in order for the
system to be bound, the smaller value of the projected separation is
necessary (see Sect.\ 2.2 and Table~\ref{table-sep}); that is, $R_p =
0.23~{\rm Mpc}$. Therefore, for the simulations, we assume that the
center of the two groups of particles are separated by $0.23~{\rm
Mpc}$ in the transverse $x$ direction. This implies that the
separation in the $z$ direction is $0.23 \times {\rm tan} \alpha$ (see
Figure~\ref{schem}).  For simplicity, we assume an $\alpha$ value of
$45^o$ degrees (see Table~\ref{table-sol}). We also assume that the
groups are separated only in the $(x,z)$ plane, i.e. there is no
displacement in the $y$ direction. Each particle in the group is now
given an initial position $(x,y,z)$ around the group center.  Because
of the small number of confirmed group members, the radial profiles of
the CL0023+0423 groups are highly uncertain; therefore, for each
simulated group we randomly choose a radius which is within $\pm 50\%$
of the observed harmonic radius $R_h$ (see Table~\ref{table-mass}).
We have chosen such wide range in radius, rather than $\pm 2\sigma$ as
we have done with the group masses, because the harmonic radius may be
a significant underestimate of the typical group radius because it is
biased toward close pairs (see Paper II).  When the radius is chosen,
each particle is given a random position within a three-dimensional
sphere of this radius.

In order to calculate the initial velocity $(u,v,w)$ of each particle,
we first need to adopt a radial velocity difference ($V_r$) between
the two groups. We do this by randomly choosing a value based upon the
observed $V_r = 2922 \pm 216$. We randomly choose an initial $V_r$
which is within $\pm 2 \sigma$ of this value.  Each particle,
therefore, has an additive velocity due to relative motion between the
two groups. For the particles in Group A, this means an additional
velocity of $-0.5 V_r$ in the $z$ (line-of-sight) direction and a
velocity of $-\frac{0.5 V_r}{{\rm tan} \alpha}$ in the $x$
direction. Similarly, for the particles in Group B, this velocity is
$+0.5 V_r$ in the $z$ direction and $+\frac{0.5 V_r}{{\rm tan}
\alpha}$ in the $x$ direction (see Figure~\ref{schem}).  We assume no
additive velocity in the $y$ direction. In addition to this relative
velocity, each particle has a velocity due to its internal motion
within the group. This motion is the result of the group's velocity
dispersion. Therefore, we first randomly choose a velocity dispersion
that is within $\pm 2 \sigma$ of the observed values (see
Table~\ref{table-mass}).  Each particle in the group is then given an
extra velocity in each dimension $(u,v,w)$ based on this dispersion,
assuming that the velocities within the group are isotropic and have a
Gaussian distribution (see Figure~\ref{velhist}).

We have now specified all the initial conditions for an individual
simulation. In addition, the total number of timesteps, the duration
of a timestep, the accuracy parameter $\theta$, and the softening
length $\epsilon$ must be specified. The accuracy parameter determines
the size of the largest cell, in units of cell distance, which is not
subdivided further (Barnes \& Hut 1986,1989). The softening parameter
is defined such that the maximum force between two particles occurs at
distance of $\frac{\epsilon}{\sqrt{2}}$ (see e.g.\ Binney \& Tremaine
1987).  We run each simulation for $4 \times 10^{9}$ yrs with
timesteps of $5 \times 10^{6}~{\rm yrs}$.  Based on empirical tests of
our own and the error analysis of Barnes \& Hut (1989), we have chosen
an accuracy parameter of $\theta = 1.0$ and a softening parameter of
$\epsilon = 1~{\rm kpc}$. With these parameters energy is conserved to
better than 5\%.

In order to perform a statistical analysis of the results, we have run
1000 individual simulations.  We consider a final system to be
``merged'' if two criteria are satisfied at the final timestep of $t =
4~{\rm Gyr}$ (or $z \approx 0$).  The median position of the particles
that were originally in Group A and B are calculated.  The separation
between these two positions must be less than 4 Mpc.  Secondly, their
median radial velocities are calculated. The final velocity difference
between the Group A and Group B particles must be less than the
half-width of velocities in the entire system.  Based on these
criteria, we find that the two groups of particles have merged in 201
out of 1000 simulations.  We have examined the final configuration for
each of these simulations. The adopted merger criteria is quite
conservative. All systems visually appear merged in the observable
$(x,y)$ plane, as well as in the radial velocity distribution. A
sample simulation which has successfully merged is shown in
Figure~\ref{samp-merg}.  The radial velocity dispersion of the final
system of particles is $\sim 785~{\rm km~s^{-1}}$, typical of an Abell
richness class 1 cluster. A sample simulation which has not merged is
shown in Figure~\ref{samp-nomerg}.  It is clear from the resulting
figure that this system is still separated in both radial velocity and
position on the sky.  The results of the simulations are fairly
insensitive to the total number of particles or the number of
particles per group. We have rerun the simulations with varying total
particle number and ratio of particles in Group B to Group A (B:A
always less than 1).  We find that in all cases $15 \pm 8\%$ of the
simulations satisfy our criteria for a merger, consistent with the
percentage found above.

In Figure~\ref{merg-stats}, we show the distributions of initial
parameters ($M$, $\sigma$ and $R_h$) for Group A and B in those
simulations where the two groups successfully merge. The median value
(and the deviation as determined by the interquartile range) for each
parameter is given in Table~\ref{table-mstats}. The distributions of
the parameters for the less massive system, Group A, are almost flat,
indicating that the simulations are relatively insensitive to this
system; however, there is a clear trend in the mass $M$ of the more
massive system, Group B. This distribution is much more heavily
weighted to larger masses.  These results are consistent with the
analytic analysis of Sect.\ 3 where we have also shown that more mass
is required to bind this system. The median value of the mass of Group
B is $6.63 \times 10^{14}~{\rm M_{\odot}}$, or approximately 1$\sigma$
more than the observed mass of CL0023B.  In addition, in
Figure~\ref{merg-vr} we show the distribution of the radial velocity
separation ($V_r$) for those simulations which have produced
mergers. As expected, this distribution favors lower values of $V_r$
with the median value of $2642 \pm 165~{\rm km~s^{-1}}$.

We can now examine the properties of the final systems of particles in
those simulations where the two initial groups have merged.  The final
system of particles has properties which are typical of larger
associations of galaxies, i.e.\ a cluster of galaxies. In
Figure~\ref{cl-stats}, we show the distributions of the velocity
dispersion ($\sigma$) and the harmonic radius $R_h$ in the merged
systems at the final timestep of 4 Gyr.  For both calculations, we
have used only those particles which have a radial velocity difference
relative to the mean cluster velocity of less than $3000~{\rm
km~s^{-1}}$. The median radial velocity dispersion is $729 \pm 63~{\rm
km~s^{-1}}$; these values are consistent with an Abell Richness class
1 cluster (Dressler 1980a; Bahcall 1981; Struble \& Rood 1991).  The
median harmonic radius is $0.48 \pm 0.06~{\rm Mpc}$.  The values of
$R_h$ are consistent with those found in local clusters where $R_h$ is
typically less than 1.0 Mpc (e.g.\ Beers, Geller \& Huchra 1982;
Chapman, Geller \& Huchra 1987; Postman, Geller \& Huchra 1988).
After 4 Gyrs, the cluster has had sufficient time to relax and form a
central, compact core.  During the first few Gyrs, the two groups make
several passes through each other, finally merging at the second or
third encounter (see also Nakamura, Hattori \& Mineshige 1995). During
this period, the cluster appears more analogous to local irregular or
open clusters (such as Virgo) which exhibit a more uniform galaxy
overdensity and a higher degree of irregularity (e.g.\ Abell 1958;
Dressler 1980a,b; Abell, Corwin \& Olowin 1989). All of the merged
systems already appear as clusters $\sim 1$ Gyr after the start of the
simulations, though the velocities of the system particles are not yet
completely randomized. At the redshift of CL0023+0423 ($z = 0.84$),
this implies that these systems will be observable clusters at
redshifts of $z \simless 0.5$.

In summary, we have shown that, within the observational
uncertainties, the group-group system of CL0023+0423 has enough mass
and is close enough in both positional and velocity space to merge
into a verifiable cluster of galaxies. If these two groups do merge,
the resulting system will appear as a cluster of galaxies within $\sim
1$ Gyr or at a redshift of $z \sim 0.5$. Because the system is
relatively young, noticeable dynamic evolution within the cluster is
still taking place after 3 Gyr ($z \simless 0.1$).

\section{Discussion}

In this paper, we have used an analytic analysis, as well as N-body
simulations, to examine whether the observed group-group system of
CL0023+0423 will eventually merge. Both studies have shown that this
system is most likely unbound and simply a chance projection on the
sky.  However, within the observational uncertainties, there do exist
some system configurations where the two groups are bound and are in
the process of merging. In the analytic analysis of the system
dynamics, we have shown that there are bound-incoming solutions for
this two body system if the radial separation of the two groups is on
the order of $\sim 0.23~{\rm Mpc}$, the total mass of the system is
larger by 1$\sigma$ ($M \sim 7 \times 10^{14}~h^{-1}~{\rm
M_{\odot}}$), and the radial velocity difference between the two
groups is smaller by 1$\sigma$ ($V_r \sim 2700~{\rm km~s^{-1}}$).  In
addition, we have run 1000 N-body simulations with initial parameters
that are randomly chosen to be within $\pm 2\sigma$ of the observed
values.  We find that the two groups have successfully merged in 201
simulations.  Therefore, if the formal errors are an accurate measure
of the uncertainty in the observational parameters, the group-group
system of CL0023+0423 has an 20\% chance of merging within a few Gyrs.
This probability is, however, optimistic as the successful mergers
depend strongly on certain parameters, specifically the mass of
CL0023B and the radial velocity difference between the two groups.
The N-body simulations indicate that, for a merger to occur, the
system must have a mass which is on the high end and a radial velocity
difference which is on the low end of the possible range in observed
values. However, if the groups are not truly virialized, the mass will
be overestimated by as much as a factor of 2 (Small et al.\ 1998);
therefore, the choice of a large mass would go in the opposite
direction of the likely systematic bias.  If the two groups eventually
merge, the resulting system has a characteristic velocity dispersion
and harmonic radius which are typical of a richness class 1 or 2
cluster.

For the discussion of the implications for cluster formation, we will
assume that the group-group system in CL0023+0423 will eventually
merge. The simulations indicate that the merged systems will appear as
a cluster of galaxies within $\sim 1-2~{\rm Gyr}$ (at $z \simless
0.5$); therefore, the groups of galaxies in CL0023+0423 are likely to
be the progenitors of some intermediate-redshift clusters.  As
discussed in Sect.\ 4, the resulting cluster is still evolving
dynamically after 3 Gyrs or at redshifts of $z \simless 0.1$.
Therefore, over this time period, the system would appear more like an
open cluster which are irregular, loose, and presumably dynamically
young.  Several clusters belonging to this class have been studied at
redshifts of $z = 0.31 - 0.54$.  High-resolution HST imagery of these
clusters have been used to measure the morphological composition of
the galaxy populations.  Because the number of spectroscopically
confirmed cluster members is limited, the morphological fractions are
determined statistically using background-corrected distributions of
morphologies.  These studies imply elliptical fractions between
$27-47\%$ and total early-type (E + S0) fractions between $45-80\%$
(Dressler et al.\ 1997; Stanford, Eisenhardt \& Dickinson 1997;
Andreon, Davoust \& Heim 1997; Couch et al.\ 1998).  Classifiers of
these galaxies estimate that the errors in a particular morphological
bin is approximately $\pm 20\%$ (see Smail et al.\ 1997; Andreon,
Davoust \& Heim 1997; Andreon 1998). Comparisons between independent
studies confirm these uncertainties and may imply even larger ones;
for example, take the open cluster CL0939+4713 at a redshift of $z =
0.41$. Andreon, Davoust \& Heim (1997) find an elliptical fraction of
15\%, while Dressler et al.\ (1997) find 37\% ($M_V < -20; h = 1, q_0
= 0.5$). These fractions are statistically different and are not the
result of small number statistics (Andreon 1998).  The fractions for
the total early-type population appear more consistent. For this
cluster, independent classifiers find early-type fractions of 49, 56,
and 60\%, respectively (Andreon, Davoust \& Heim 1997; Dressler et
al.\ 1997; Stanford, Eisenhardt \& Dickinson 1997). However, as
implied by the large discrepency in the elliptical fraction, there is
contention about the ratio of elliptical to S0 galaxies in both open
and compact clusters at these redshifts.  Dressler et al.\ (1997) have
examined 10 clusters at $z = 0.37 - 0.56$ and find that, compared to
local clusters, the fraction of S0 galaxies is smaller by a factor of
$\sim 2-3$. However, other classifiers of the same clusters do not
find this discrepency (Stanford, Eisenhardt \& Dickinson 1997;
Andreon, Davoust \& Heim 1997). Therefore, it is fair to say that the
exact morphological fractions in intermediate-redshift clusters are
still not accurately known.

In Paper III, we have performed a similar morphological analysis on
the two groups in CL0023+0423 using both high-resolution HST imagery
and the spectral characteristics of the member galaxies.  The
morphological classifications of the galaxies within the HST
field-of-view ($\sim 150^{''} \times 150^{''}$) were made according to
the Revised Hubble system of nearby galaxies (e.g.\ Sandage 1961;
Sandage \& Bedke 1994).  The spectroscopic coverage of the HST field
is limited with only 12 confirmed members. Seven are classified as
spiral or irregular galaxies. The remaining five have been classified
as either an elliptical or compact galaxy; none were classified as S0
galaxies.  We expect that these identifications are relatively robust
and directly comparable to the classifications of nearby galaxies as
we are morphologically classifying these high-redshift galaxies just
slightly bluewards of the well-studied $B$ band and to roughly the
same limiting surface brightness as local samples (see Paper III).  Of
the five galaxies classified as ellipticals, three show old, red
absorption spectra typical of present-epoch ellipticals. The other two
have asymmetric disks and are identified as possible mergers (see
Paper III).  Their spectra show strong star-forming features and are
consequently bluer than the other ellipticals. Their properties are
more consistent with those of blue compact galaxies found at redshifts
of $z \sim 0.1 - 0.7$ in previous HST observations (Koo et al.\ 1994,
1995). All other galaxies in the HST field-of-view with strong
\ion{O}{2} emission (equivalent widths of greater than 10 \AA) are
classified as spirals or irregular/peculiar galaxies.  Therefore,
based on both spectral and morphological evidence, we find an
early-type fraction of 25\% (3 out of 12) in the central region of the
CL0023+0423 system.  If we try to improve the statistics by examining
the spectral characteristics of the confirmed group members over the
larger LRIS field-of-view ($\sim 2^{'} \times 8^{'}$; see Paper I), we
find that only 17\% (4 out of 24) of the confirmed members have old,
red absorption spectra (ages of $\simgreat 3$ Gyrs; see Paper II).
Two-thirds of all group members have active star formation with
\ion{O}{2} equivalent widths which are typically much larger than 10
\AA; the average \ion{O}{2} equivalent width is 35 \AA.  If we assume
that all of these galaxies are spirals or irregulars (as is the case
for those galaxies with HST morphologies; see above), the spectral
results imply early-type fractions of 33\% or less.  

We have examined the effect of sample completeness on these
percentages, i.e.\ the possibility that some of the galaxies in this
field which have spectra but no redshift measurement are actually
system members with an absorption spectra that is consistent with a
very faint, early-type galaxy.  Based on the numbers listed in Table 1
of Paper II, we present a statistical argument to estimate the number
of such galaxies. We use the fraction of galaxies in this field that
have an absorption spectrum and are also confirmed system members
(25\%), as well as the number of galaxies down to our limiting
magnitude which have spectra but no redshift measurement (14).  If we
assume that all 14 of these galaxies have faint absorption spectra, we
expect that four (or less) are system members which have been
missed. This implies that purely absorption spectra galaxies comprise
less than 29\% of the system population. Therefore, we conclude that
our early-type fractions as estimated spectroscopically should not be
significantly biased by incompleteness.

Because the number of confirmed group members with HST morphologies is
small, we have also examined the background-subtracted distribution of
galaxy morphologies in this system.  This study indicates that
CL0023+0423 has a galaxy population which is more typical of the
field.  The numbers are consistent with 100\% of the galaxies in this
two group system being normal spirals.  Of all the galaxies brighter
than $M_V = -19.0 + 5~{\rm log}~h$ in the central $\sim
0.5~h^{-1}~{\rm Mpc}$ of this system, early-type galaxies may comprise
only $5^{+35}_{-5}\%$ of the total population.  The formal
uncertainties are large because of the small number statistics (see
Paper III); however, there are consistent with the spectral results.
If these groups are in the process of forming a cluster, a comparison
between the morphological fractions in the CL0023+0423 system and
those in open, intermediate-redshift clusters may indicate that a
non-negligible fraction of early-type galaxies are formed between
redshifts of $z \sim 0.8$ and $z \sim 0.3$.  These results are only
suggestive because of the large uncertainties in the morphological
fractions at both intermediate and high redshift.

If formation of early-type galaxies is occurring during this time
period, we would expect that a significant fraction of these galaxies
in open clusters would have spectral features which are characteristic
of star formation activity within the last $\sim 1$ Gyr.  Preliminary
spectral studies indicate that early-type galaxies in
intermediate-redshift clusters (both open and compact) show no signs
of current or recent star formation; however, there is a
non-negligible fraction which show post-starburst spectral features
(Dressler et al.\ 1994; Barger et al.\ 1996; Oemler, Dressler \&
Butcher 1997; Poggianti 1997; Couch et al.\ 1998).  Starbursts are
thought to be the result of mergers or interactions between
galaxies. Such events can drive substantial evolution in not only the
stellar population, but also the morphological appearance of the
galaxy (e.g.\ Schweizer 1986; Hibbard et al.\ 1994). The morphological
signatures of a merger between two disk galaxies, such a long, bright
tidal tail, would be visible for only $\sim 1-2$ Gyrs.  After this
time, the tidal debris becomes indistinguishable from the main merger
remnant, a morphologically normal elliptical galaxy (Mihos 1995,
1998).  These timescales are consistent with the transformation of
disk galaxies to spheroids between redshifts of $z \sim 0.8$ and $z
\sim 0.5$.  In fact, the fraction of galaxies in intermediate-redshift
clusters which exhibit signs of merging events, such as distorted
morphologies or tidal tails, is as large as $\sim$20\% (e.g.  Lavery
\& Henry 1988; Lavery, Pierce \& McClure 1992; Dressler et al.\ 1994;
Barger et al.\ 1996; Oemler, Dressler \& Butcher 1997; Couch et al.\
1998).  In the CL0023+0423 system, we also find that 41\% (5 out of
12) of the confirmed cluster members with HST morphologies have either
a ring structure, an asymmetric disk, or a double nuclei which may
indicate an interaction or merger.  In addition, Koo et al.\ (1995)
suggest that blue compact galaxies (of which there are two in this
system; see above) are galaxies which have undergone a recent, strong
burst of star formation.  The galaxy luminosity will eventually fade
by several magnitudes, resulting in a system with a surface brightness
and velocity width which are typical of nearby low-luminosity
spheroids (see Kormendy \& Bender 1993). The wide range in redshift
($z \sim 0.1 - 0.85$) over which these galaxies are found also implies
that major star formation episodes have occurred in some spheroids
over many Gyrs.  Because the group velocity dispersion is on the order
of the internal velocity of its member galaxies, the kinematics of
poor groups makes them a favorable environment for galaxy-galaxy
encounters (Aarseth \& Fall 1980; Barnes 1985; Merritt 1985).
Therefore, it is conceivable that groups of galaxies like those of
CL0023+0423 may be the sites of morphological evolution in galaxies.

Signatures of past starburst activity are not uncommon in local
environments. For example, studies of nearby rich clusters with
substantial substructure, as well as local groups of galaxies,
indicate that $\sim 15\%$ of the early-type member galaxies have
experienced recent or ongoing star formation (Caldwell \& Rose 1997;
Zabludoff \& Mulchaey 1998). Field ellipticals also show remnants of
past violent activity. Line strengths indicate that approximately 25\%
of all field ellipticals have produced 10\% or more of their stars
since redshifts of $z \sim 0.5$ (Trager 1997).  These fractions are
even higher for S0 galaxies where $25-50\%$ show such activity
(Fisher, Franx \& Illingworth 1996; Trager 1998). These results also
suggest that some early-type galaxies have experienced and are still
experiencing considerable evolution since $z \sim 1$ (see also
Schweizer 1997). 

If the results of the study of CL0023+0423 do support early-type
formation at relatively recent epochs, it does not preclude in any way
the formation of these galaxies at very high redshift or,
subsequently, the existence of large populations of early-type
galaxies in high-redshift clusters.  In fact, much data, including
galaxy age and color scatter, support a very early epoch ($z > 5$) for
the formation of early-type galaxies (e.g.\ Dickinson 1995; Steidel et
al.\ 1996; Ellis et al.\ 1997; Oke, Gunn, Hoessel 1997; Stanford,
Eisenhardt \& Dickinson 1997; Paper II). Additionally, already
well-formed clusters at $z > 0.8$ appear to have fractions of
early-type galaxies which are comparable to local clusters (Stanford,
Eisenhardt \& Dickinson 1997; Stanford et al.\ 1997; Dickinson 1997;
Lubin et al.\ 1998). The relaxed, well-formed nature of these systems
suggest that they formed at a much earlier epoch, allowing for perhaps
enough time to create a significant population of early-type galaxies.
The observations of the CL0023+0423 system may simply imply that a
fraction of the early-type population is forming and/or undergoing
significant activity at redshifts of $z < 1$.

On the other hand, the fact that a sizable portion of the early-type
population in certain clusters are formed at redshifts of $z < 1$ may
be difficult to reconcile with the tight color-magnitude relation
found in clusters of galaxies between redshifts of $z = 0$ and $z \sim
0.9$ (e.g.\ Stanford, Eisenhardt \& Dickinson 1995, 1997; Ellis et
al.\ 1997). The small intrinsic color scatter in early-type galaxies
implies that the fractional differences in age between the stellar
population in these galaxies are small (Bower, Lucey \& Ellis 1992).
Because the color scatter in the E/S0 populations remains roughly
constant to $z \sim 0.9$, this suggests that the last period of major
star-formation took place at redshifts significantly greater than 1
(see also Arag\'on-Salamanca et al.\ 1993; Stanford, Eisenhardt \&
Dickinson 1995; Paper II). The majority of the clusters in these
studies have been centrally-concentrated, compact clusters. Cluster
concentration, which correlates well with degree of regularity, is
measured by the compactness parameter $C \equiv {\rm
log}(R_{60}/R_{20})$ where $R_{60}$ and $R_{20}$ refer to the radii
containing 60\% and 20\% of the cluster population, respectively
(Butcher \& Oemler 1978).  It is conceivable that open, irregular
clusters ($C < 0.35$) have systematically larger E/S0 color scatters
which are the result of their more recent evolutionary history.  Five
open clusters have been well-studied at intermediate redshifts,
CL0303+17, CL0412-65, CL0939+47, CL1447+23, and CL1601+42.  The color
scatter in their early-type populations are all consistent with the
mean scatter of the entire cluster sample (Ellis et al.\ 1997;
Stanford, Eisenhardt \& Dickinson 1997).  However, the relatively
large uncertainties in the estimated scatters may still allow for a
systematic difference between open and compact clusters. There are
other apparent differences between these types of clusters; for
example, the morphology-density relation which is characteristic of
both open and compact clusters locally is qualitatively similar in
compact clusters at intermediate redshift but completely absent in the
loose, open clusters (Dressler et al.\ 1997).

Even if the two CL0023+0423 groups are not in the process of merging,
the observations of this system may have implications for the
evolution of groups of galaxies.  CL0023A and CL0023B have similar
kinematic characteristics to local groups.  Their velocity dispersions
and estimated masses are typical of well-studied groups in the nearby
universe (Hickson 1982; Ramella, Geller \& Huchra 1989; Mulchaey et
al.\ 1996; Zabludoff \& Mulchaey 1998). Locally, there exists a
well-established correlation between the early-type fraction and the
group velocity dispersion (Hickson, Kindl \& Huchra 1988; Zabludoff \&
Mulchaey 1998).  As the velocity dispersion increases from 100 to
$450~{\rm km~s^{-1}}$, the early-type fraction rises from effectively
zero to 55\%.  These calculations include all member galaxies which
are brighter than the absolute magnitude limit of $M_B \sim -16$ to
$-17 + 5~{\rm log}~h$ and within a radius of $\sim 0.6-0.8~h^{-1}~{\rm
Mpc}$.  Systems with larger velocity dispersions move into the more
massive cluster regime.  Here, the relation between early-type
fraction and velocity dispersion flattens out with clusters of
galaxies typically having an early-type fraction of $\sim 55 - 65\%$
within these magnitude and radial limits (Dressler et al.\ 1980a;
Whitmore, Gilmore \& Jones 1993).  Most clusters at $z \simgreat 0.1$
are not sampled to such faint limits. If a brighter magnitude limit
were adopted on the local group data ($M_B \sim -18 + 5~{\rm log}~h$),
the statistics worsens though the early-type fraction does increase as
the brightest galaxies in groups and clusters are normally
early-types.  For the groups with the highest velocity dispersion, the
early-type fraction approaches 80\% or more, again consistent with the
morphological composition of present-day clusters (Oemler 1974;
Dressler et al.\ 1980a; Whitmore, Gilmore \& Jones 1993; Dressler et
al.\ 1997).

Based on the local correlation between early-type fraction and
velocity dispersion described above, the modest early-type fractions
in both CL0023+0423 groups may imply that the galaxy populations in
groups of galaxies have evolved (see also Paper III).  We explore this
possibility by examining the spectroscopically-confirmed group
members. These data corresponds to an absolute magnitude limit of $M_B
\sim -19$ to $-20 + 5~{\rm log}~h$ (see Paper II).  In CL0023A, four
of the 7 galaxies are morphologically classified as spirals (see Paper
III). Of the remaining three galaxies, one is classified as an
elliptical. The other two are not within the HST field-of-view and,
therefore, have no direct morphological information; however, they
both have very high equivalent width \ion{O}{2} emission (20 and 28.6
\AA, respectively). Based on the spectroscopic and morphological data
from the two cluster fields presented in Papers II and III, we find
that 86\% of all galaxies with such strong \ion{O}{2} emission are
classified as either a spiral or irregular/peculiar; the other 14\%
are classified as ellipticals though their photometric and spectral
properties indicate that they are more similar to blue compact
galaxies (see above).  This correlation between morphology and
\ion{O}{2} emission would suggest that the remaining two galaxies in
CL0023A are most likely late-type galaxies.  Therefore, the elliptical
fraction would be only 14\% (1 out of 7).  Indeed, only 29\% (2 out of
7) of the confirmed members have a typical K star absorption spectra
which is old and red. These two galaxies are classified as an
elliptical (MDS ID \#17; see Sect.\ 5.3.1 of Paper III) and a peculiar
Sa (MDS ID \#82; see Sect.\ 5.3.2 of Paper III).  Based on the local
relation, we would expect an early-type fraction of $10^{+20}_{-10}\%$
for this low dispersion system of $158^{+42}_{-33}~{\rm km~s^{-1}}$.
The expected fraction should actually be higher, perhaps 33\%, as the
local data goes over two magnitudes fainter. In addition, with only 7
redshifts, the velocity dispersion of this group is likely biased low
by a factor of $\sim 1.5$ (Zabludoff \& Mulchaey 1998).  This bias
would also imply a predicted early-type fraction which is higher.

CL0023B has 17 confirmed members, of which 7 have direct morphological
information. Of these seven, three are spirals, two are blue compact
galaxies (classified as ellipticals), one is an elliptical, and one is
classified as compact (see Paper III). Only 2 of these 7 galaxies have
both spectroscopic and morphological properties which are
characteristic of an elliptical galaxy.  Overall, 11 out of 17
confirmed group members have extremely strong \ion{O}{2} emission (an
average equivalent width of 43 \AA). If we assume that all of these
are late-type galaxies as is true for the vast majority of galaxies
with direct morphological information (see above), this suggests an
early-type fraction of 35\% or less. In fact, only 12\% (2 out of 17)
of the confirmed group members have an old, red K star absorption
spectrum.  These two galaxies are morphologically classified as an
elliptical (MDS ID \#20) and a compact galaxy (MDS ID \#115; see
Sect.\ 5.3.1 of Paper III). For CL0023B with a velocity dispersion of
$415^{+102}_{-63}~{\rm km~s^{-1}}$, the local relation would predict
an early-type fraction of $55^{+10}_{-20}\%$.  For the brighter
magnitude cut which is characteristic of our high-redshift data, this
fraction would be 80\% or more.

These observations suggest that there is no strong correlation between
early-type fraction and velocity dispersion in the CL0023+0423 system.
Consequently, the difference between our observations and that
predicted by the local relation may imply early-type formation and,
therefore, a change in the morphological composition of groups of
galaxies between redshifts of $z \sim 0.8$ and the present epoch.
This evolution would suggest that, in order to create the correlation
between velocity dispersion and early-type fraction that exists in
local groups of galaxies, the new early-type galaxies will form only
in the more massive systems.  These hypotheses are only valid if we
assume that the groups of CL0023+0423 are typical of all groups at
high redshift and that high-redshift groups are the progenitors of
local galaxy groups.  It is also important to stress that the very
limited number of confirmed members with direct morphology imply large
statistical (and perhaps systematic) errors in both the velocity
dispersion and early-type fractions of both components.  Therefore, we
cannot say for certain that the properties of the CL0023+0423 groups
are truly inconsistent with the early-type fraction versus velocity
dispersion relation observed locally.  Indeed, even the nearby
relation contains a large scatter with many local groups being
significant outliers from the best-fit correlation (Zabludoff \&
Mulchaey 1998).

The observations of CL0023+0423 have provided us with the first
detailed data on the optically-selected groups of galaxies and a
potential group-group merger at high redshift. Though broad
conclusions certainly cannot be made based on only two high-redshift
groups, the results are suggestive, implying that there may be
substantial evolution in galaxy morphology occurring at redshifts of
$z < 1$.

\vskip 0.5cm

We would like to thank the referee Ann Zabludoff for very useful
comments on the text.  We also thank J.S. Mulchaey and S.C. Trager for
helpful discussions and P. Brieu and L. Hernquist for expert advice on
the N-body simulations.  The W.M. Keck Observatory is operated as a
scientific partnership between the California Institute of Technology,
the University of California, and the National Aeronautics and Space
Administration.  It was made possible by generous financial support of
the W. M. Keck Foundation.  Support for LML was provided by NASA
through Hubble Fellowship grant HF-01095.01-97A awarded by the Space
Telescope Science Institute, which is operated by the Association of
Universities for Research in Astronomy, Inc., for NASA under contract
NAS 5-26555.  This research was supported in part by {\it HST} GO
analysis funds provided through STScI grant GO-06000.01-94A.



\newpage

\textheight=9.0in
\textwidth=7.0in
\voffset -0.75in
\hoffset -0.5in

\newcommand{\ra}[7]{$#1$ $#2$ $#3$ {\tt #4} $\!\!#5$ $#6$ $#7$}

\begin{deluxetable}{clclcccc}
\tablewidth{0pt}
\tablenum{1}
\tablecaption{Dynamical Parameters of CL0023+0423}
\tablehead{
\colhead{System} &
\colhead{N$_z$} &
\colhead{$\overline z$} &
\colhead{$\sigma~({\rm km~s^{-1}})$} &
\colhead{$M_{PW} \ (10^{14}~{\rm M_\odot})$} &
\colhead{$M_{PM} \ (10^{14}~{\rm M_\odot})$} &
\colhead{$M_{RW} \ (10^{14}~{\rm M_\odot})$)} &
\colhead{$R_h$ (Mpc)}}
\startdata
0023A&  7 & 0.8274&$158^{+42}_{-33}$   & $0.10^{+0.05}_{-0.04}$ & $0.36 \pm 0.05$ & $0.33 \pm 0.05$ & $0.116 \pm 0.003$  \\
0023B& 17 & 0.8453&$415^{+102}_{-63}$  & $2.60^{+1.27}_{-0.79}$ & $4.17\pm 0.68$ & $5.61 \pm 0.91$& $0.366 \pm 0.008$  \\
\enddata
\label{table-mass}
\end{deluxetable}

\begin{table}
\tablenum{2}
\begin{center}
Table 2. Separation between the Two Groups
\vskip 0.2cm
\begin{tabular}{cccccc} \hline \hline
Case 	& Center Type & 	CL0023A & CL0023B & Separation & Physical Separation \\
	&	      & RA (J2000) Dec (J2000) & RA (J2000) Dec (J2000) &(arcsec) & ($h^{-1}~{\rm Mpc}$) \\
\hline
1	& Median Position& \ra{00}{23}{53.92}{+}{04}{23}{15.8} & \ra{00}{23}{51.81}{+}{04}{22}{40.7} & 47.3 & 0.23 \\
2	& Brightest Galaxy    & \ra{00}{23}{54.46}{+}{04}{23}{38.9} & \ra{00}{23}{47.91}{+}{04}{23}{40.7} & 103.0 & 0.50 \\
\hline
\end{tabular}
\label{table-sep}
\end{center}
\end{table}

\begin{deluxetable}{cccccc}
\tablewidth{0pt}
\tablenum{3}
\tablecaption{Dynamical Parameters for the Bound Solutions from the Analytic Analysis}
\tablehead{
\colhead{$R_p$ (Mpc)} &
\colhead{$M~(10^{14}~{\rm M_\odot})$} &
\colhead{$t_o$ (Gyr)} &
\colhead{$\chi$ (degrees)\tablenotemark{a} } &
\colhead{$\alpha$ (degrees)\tablenotemark{a} } &
\colhead{$R_m$ (Mpc)\tablenotemark{a} } }
\startdata
0.23	& 6.85	& 3.98	& $303.7 - 315.7$ & $41.9 - 61.9$ & $2.16 - 2.18$ \\
0.23	& 5.94	& 7.96	& $316.8 - 322.6$ & $47.0 - 58.8$ & $3.26 - 3.27$ \\
\enddata
\tablenotetext{a}{The range of acceptable parameters are given for all 
bound-incoming solutions within $1 \sigma$ of the observed radial
velocity difference, $V_r = 2922 \pm 216$.}
\label{table-sol}
\end{deluxetable}

\begin{deluxetable}{clll}
\tablewidth{0pt}
\tablenum{4}
\tablecaption{Median Values and Dispersions\tablenotemark{a} of Initial Conditions in Successfully Merged Simulations}
\tablehead{
\colhead{Group} &
\colhead{$M~(10^{14}~{\rm M_\odot})$} &
\colhead{$\sigma$ (${\rm km~s^{-1}}$)} &
\colhead{$R_h$ (Mpc)}}
\startdata
A	& $0.34 \pm 0.08$	& $155 \pm 55$	& $0.11 \pm 0.04$\\
B	& $6.63 \pm 0.77$	& $419 \pm 145$	& $0.32 \pm 0.10$\\
\enddata
\tablenotetext{a}{Dispersions are measured using the interquartile range (IQR),
i.e.\ $\Delta = 0.741 \times {\rm IQR}$.}
\label{table-mstats}
\end{deluxetable}


\def\plottwo#1#2{\centering \leavevmode
    \epsfxsize=.5\columnwidth \epsfbox{#1} \hfil
    \epsfxsize=.5\columnwidth \epsfbox{#2}}


\textheight=8.5in
\textwidth=6.5in
\voffset 0.0in
\hoffset 0.0in

\newpage

\begin{figure}
\plottwo{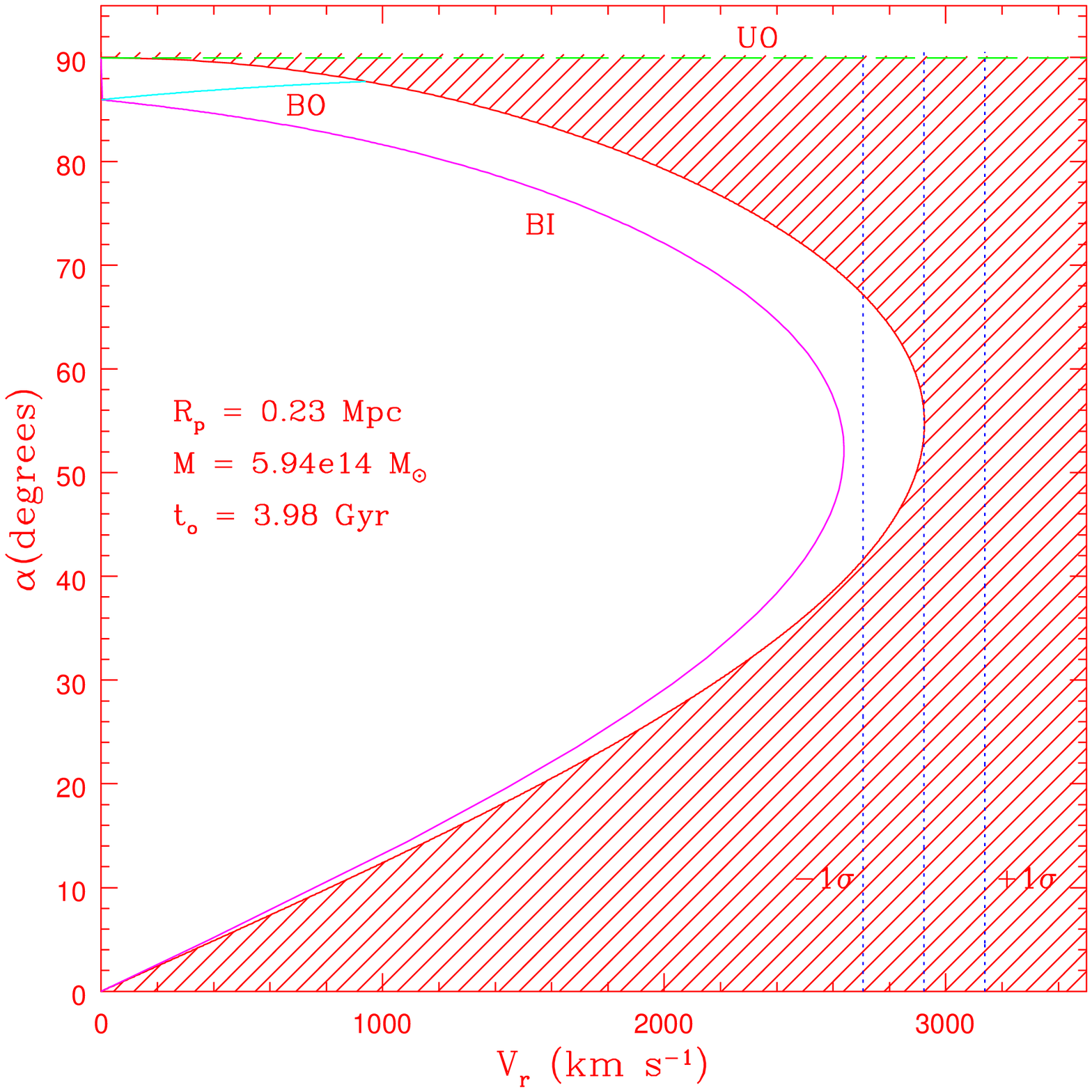}{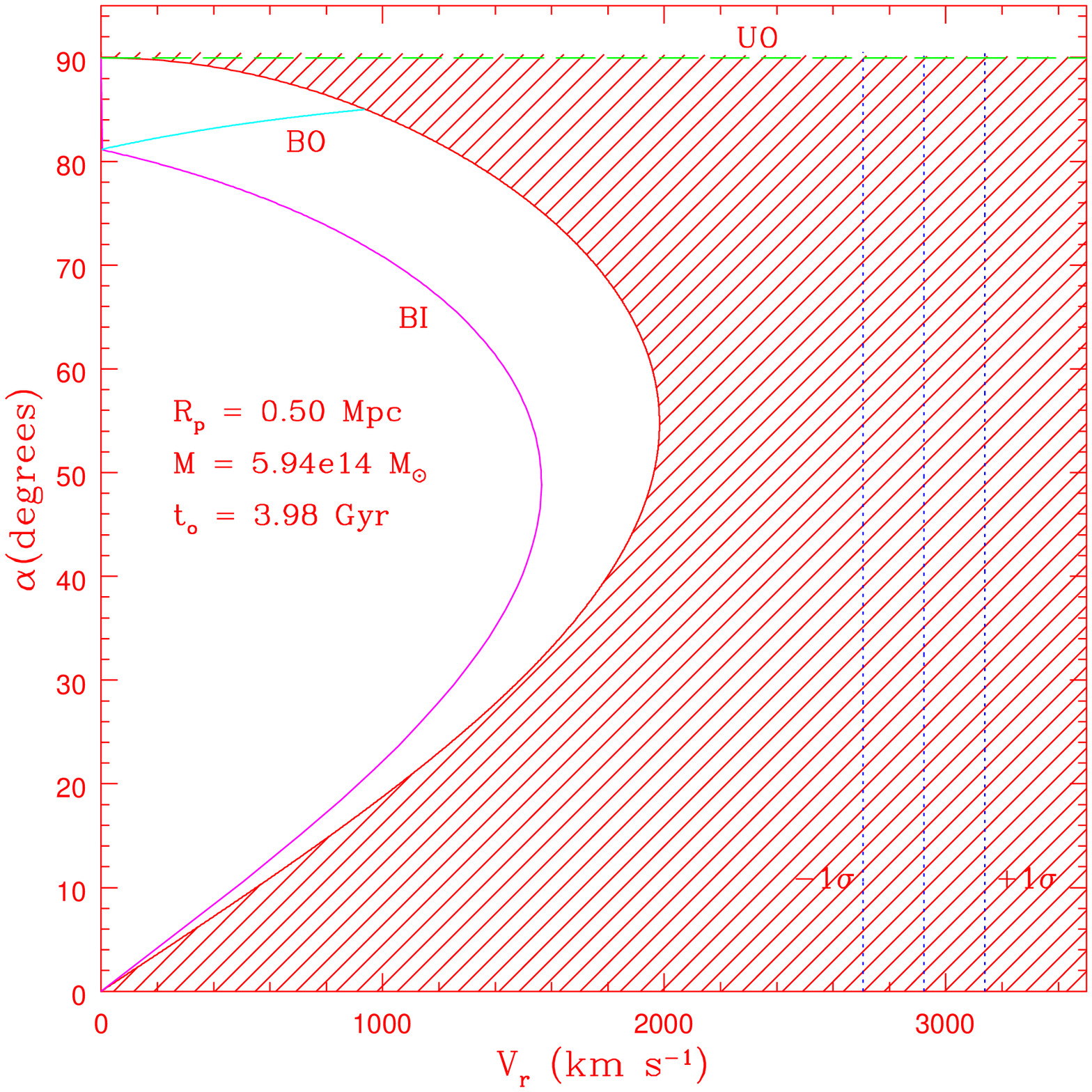}
\caption{The resulting $(\alpha, V_r)$ plane for the equations of motion
with a time $t_o = 3.98~{\rm Gyr}$ and a total system mass $M = 5.94
\times 10^{14}~M_{\odot}$. Two possible values of the projected
separation (see Sect.\ 2.2) are shown : $R_p = 0.23~{\rm Mpc}$ (left
panel) and $R_p = 0.50~{\rm Mpc}$ (right panel).  The shaded region
specifies where all unbound solutions would lie, while the unshaded
region specifies where all bound solutions would lie. The solutions
for our particular two-body problem are indicated by the solid and
dashed lines in the figure. The two solid lines indicate the bound
solutions, bound-incoming (BI) and bound-outgoing (BO); the dashed
line indicates the unbound (UO) solution. The three dotted lines at
constant $V_r$ indicate the observed radial velocity difference
between the two groups. The central dotted line is the actual value,
while the two flanking lines indicate the $\pm 1\sigma$ range. There
are no bound solutions to the equations of motion for these parameters
(within $\pm 1 \sigma$ of the observed $V_r$).}
\label{av-act}
\end{figure}

\begin{figure}
\plottwo{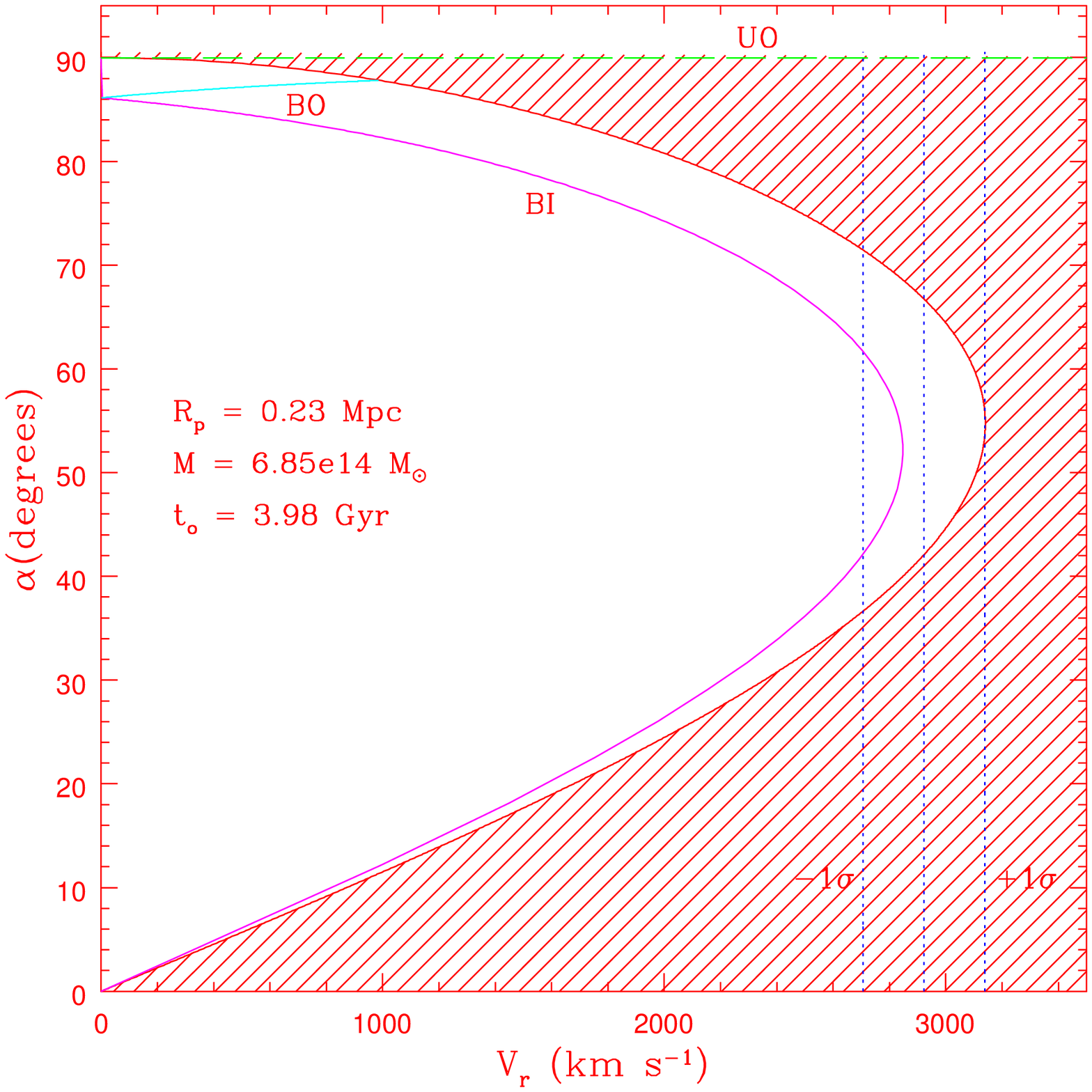}{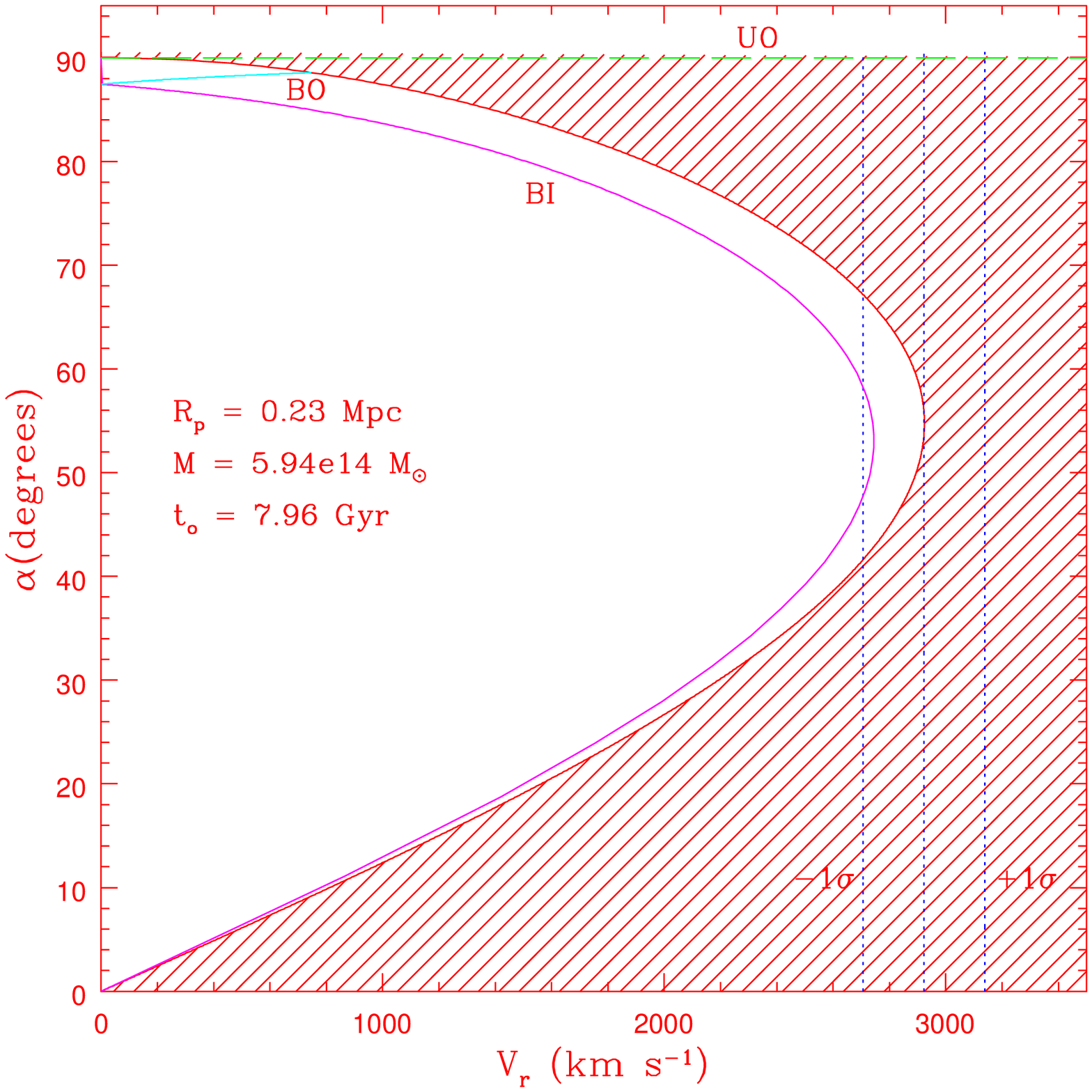}
\caption{{\it Left} : The resulting $(\alpha, V_r)$ plane for the 
equations of motion with a time $t_o = 3.98~{\rm Gyr}$ and a projected
separation $R_p = 0.23~{\rm Mpc}$. The total mass of the system is
assumed to be the observed value plus $1 \sigma$; that is, $M = 6.85
\times 10^{14}~M_{\odot}$. {\it Right} : The resulting $(\alpha, V_r)$ plane 
for the equations of motion with the observed total mass $M = 5.94
\times 10^{14}~M_{\odot}$ and a projected separation $R_p = 0.23~{\rm
Mpc}$. The time is assumed to be twice as long as that for the $q_o =
0.1$ and $H_{0} = 100~{\rm km~s^{-1}~Mpc^{-1}}$ cosmology; that is,
$t_o = 7.96~{\rm Gyr}$.  The specifications of the lines and regions
of this plot are the same as in Figure~\ref{av-act}. There are now
bound-incoming solutions within $1 \sigma$ of the observed value of
$V_r$ for a system with these parameters.}
\label{av-plus}
\end{figure}

\begin{figure}
\epsfysize=7.0in
\centerline{\epsfbox{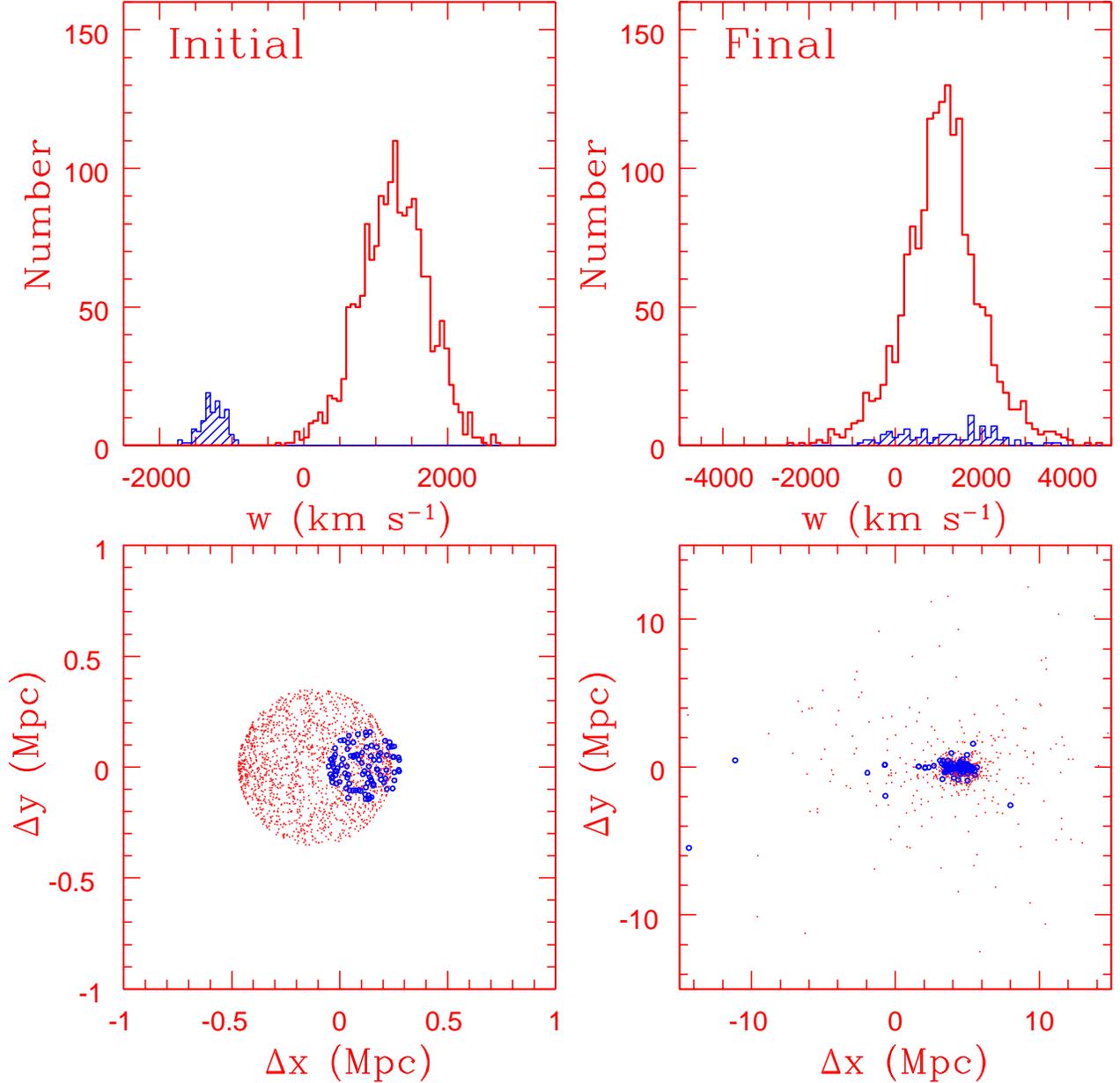}}
\caption{A sample of a simulation where the two groups have merged by
the end of the simulation. {\it Left Column} : The initial
configuration of the simulation. The upper panel shows the
distribution of the radial velocities ($w$) of the 1800 particles. The
shaded histogram represent particles from Group A, while the solid
histogram represent particles from Group B.  The bottom panel shows
the distribution of those particles on the sky, i.e.\ the $(x,y)$
plane. The particles in Group A are indicated by the open circles,
while the particles in Group B are indicated by points. {\it Right
Column} : The final configuration of the particles at $t = 4$ Gyr. The
upper panel shows the distribution of radial velocities, while the
bottom panel shows the final distribution in the plane of the sky. The
final velocity histogram indicates that the particles of Group A
clearly have a velocity distribution which is consistent with the
overall velocities of the merged system. The resulting radial velocity
dispersion of the merged system is $\sim 785~{\rm km~s^{-1}}$.}
\label{samp-merg}
\end{figure}

\begin{figure}
\epsfysize=7.0in
\centerline{\epsfbox{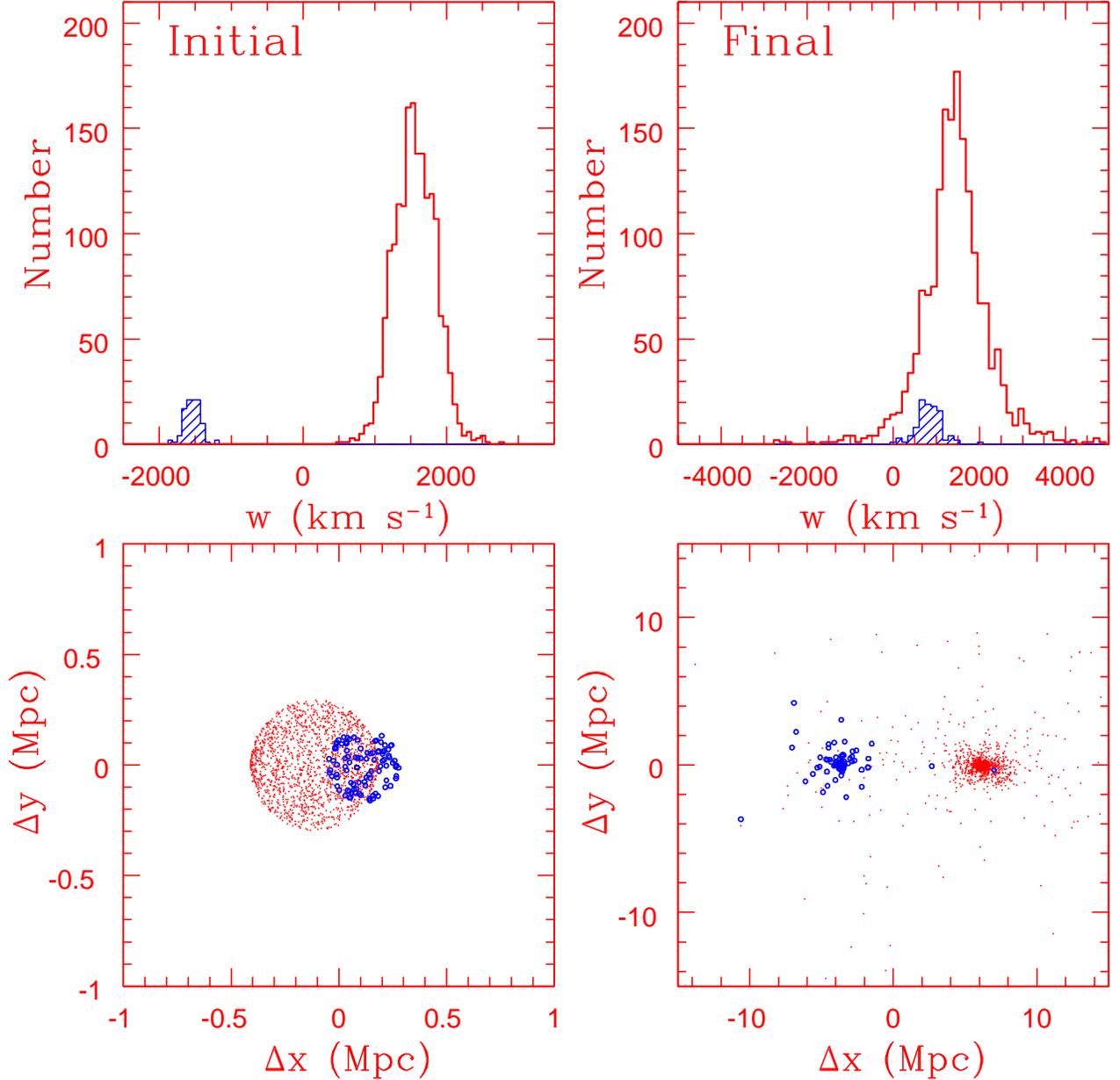}}
\caption{A sample of a simulation where the two groups have not merged by
the end of the simulation (and will never do so). {\it Left Column} :
The initial configuration of the simulation. The upper panel shows the
distribution of the radial velocities ($w$) of the 1800 particles. The
shaded histogram represent particles from Group A, while the solid
histogram represent particles from Group B.  The bottom panel shows
the distribution of those particles on the sky, i.e.\ the $(x,y)$
plane. The particles in Group A are indicated by the open circles,
while the particles in Group B are indicated by points. {\it Right
Column} : The final configuration of the particles at $t = 4$ Gyr. The
upper panel shows the distribution of radial velocities, while the
bottom panel shows the final distribution in the plane of the sky. The
final histogram of radial velocities clearly indicates that the mean
velocity of the two groups are quite different.}
\label{samp-nomerg}
\end{figure}

\begin{figure}
\epsfysize=7.0in
\centerline{\epsfbox{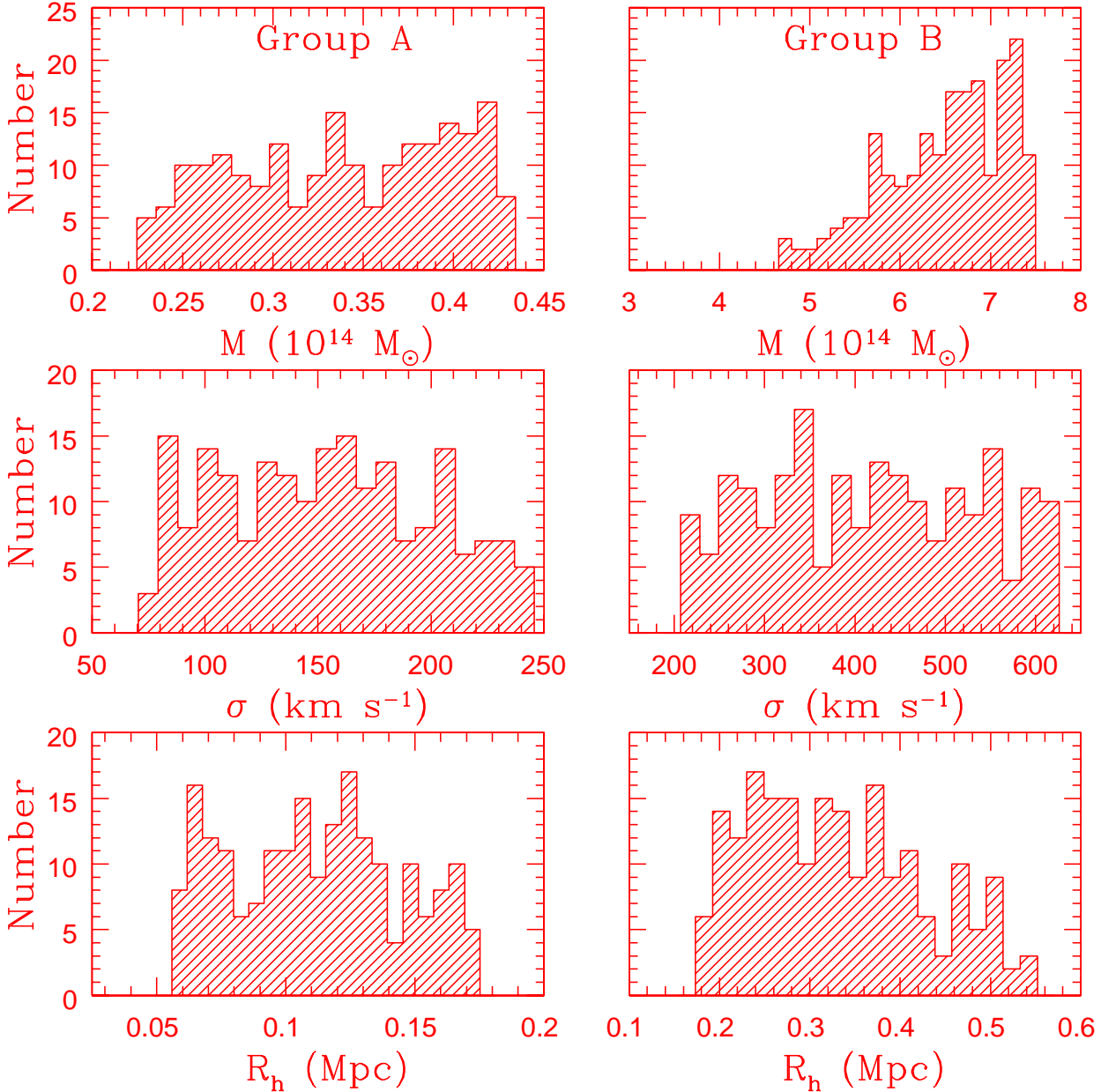}}
\caption{The distribution of the initial total mass ($M$), radial velocity
dispersion ($\sigma$), and harmonic radius ($R_h$) of Group A and B in
those simulations where the two groups eventually merged. The left
column represents those parameters for Group A, while right column
represents those parameters for Group B. It is obvious from these
distributions that the simulations are much more sensitive to the
parameters of the high mass system, Group B, specifically the initial
total mass.}
\label{merg-stats}
\end{figure}

\begin{figure}
\epsfysize=7.0in
\centerline{\epsfbox{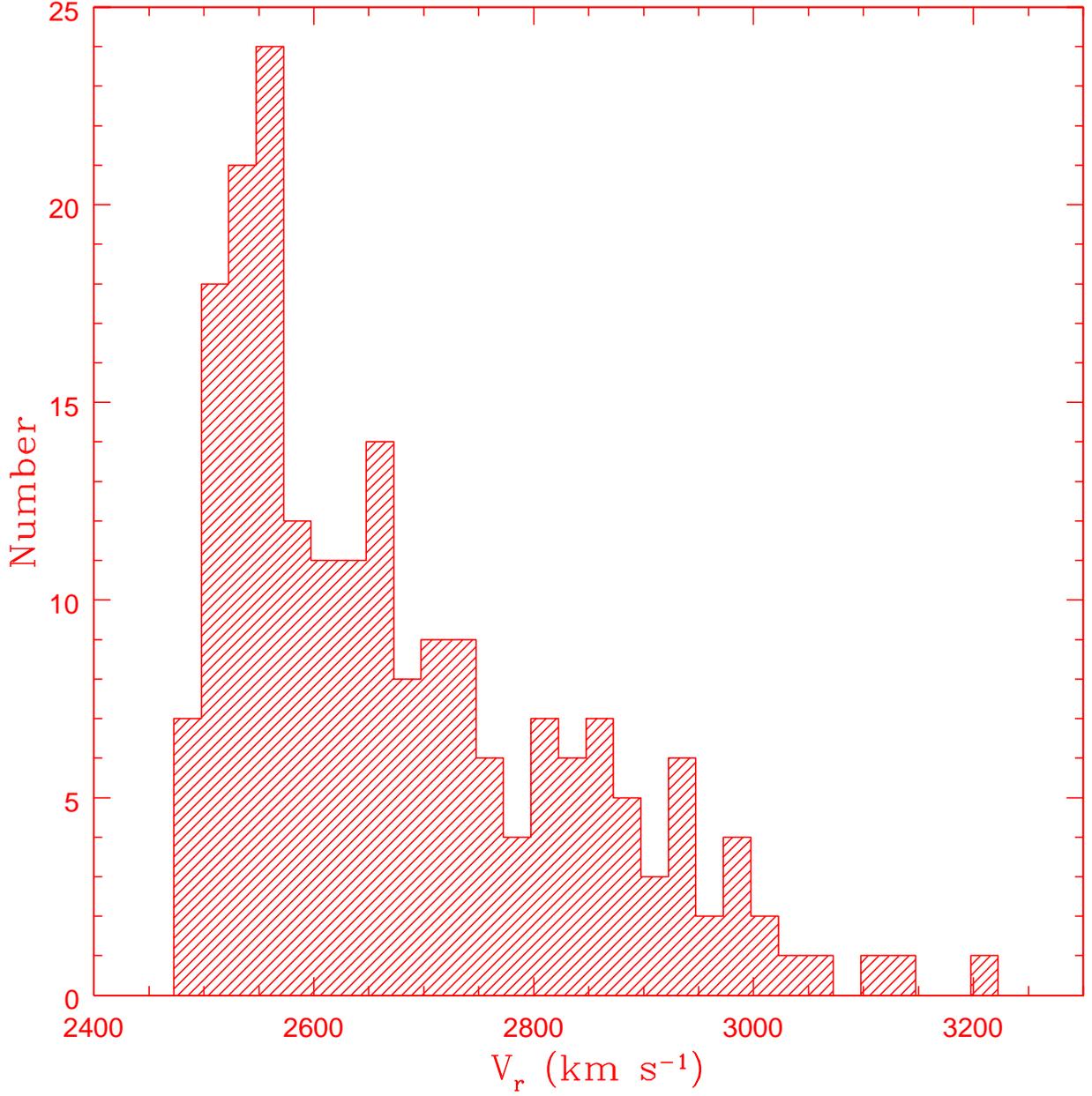}}
\caption{The distribution of the initial radial velocity
differences ($V_r$) between Group A and B for those simulations where
the two groups eventually merge. As also shown in the analytic
analysis of Sect.\ 3, the systems are more likely to merge with
smaller values of $V_r$.}
\label{merg-vr}
\end{figure}

\begin{figure}
\epsfysize=7.0in
\centerline{\epsfbox{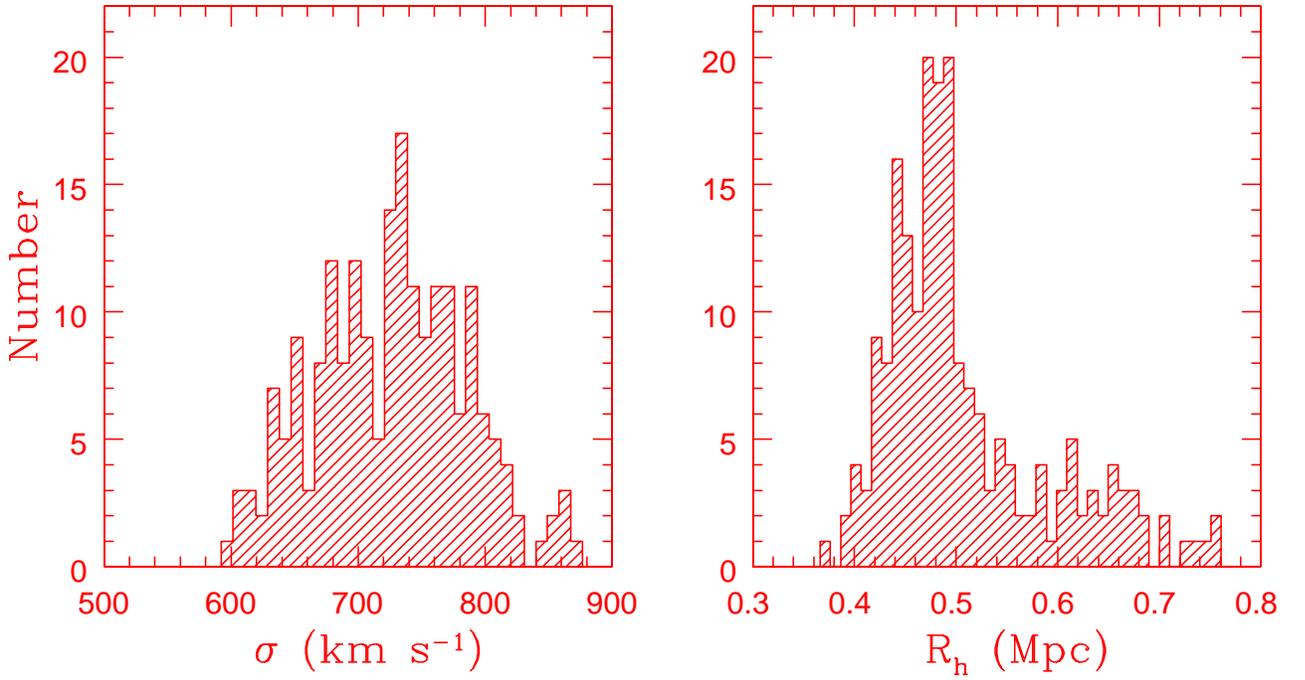}}
\caption{The distribution of the final velocity dispersions ($\sigma$)
and harmonic radii ($R_h$) of the cluster systems in those simulations
where the two groups have merged. The velocity dispersions and
harmonic radii are typical of local clusters of galaxies (see Sect.\
4).}
\label{cl-stats}
\end{figure}

\end{document}